\documentclass[twocolumn,prr,aps]{revtex4-2}

\usepackage[utf8]{inputenc}
\usepackage[T1]{fontenc}
\usepackage{amsmath}
\usepackage{braket}
\usepackage[caption=false]{subfig}
\usepackage{graphics}
\usepackage{subfig}
\usepackage{enumitem}
\usepackage{qcircuit}
\usepackage{tikz}
\usepackage[toc,page]{appendix}
\usepackage{ragged2e}

\usepackage{color}
\usepackage{epsfig}
\usepackage{latexsym}
\usepackage{amssymb}
\usepackage{amsmath}
\usepackage{wrapfig}
\usepackage[english]{babel}
\usepackage{times}
\usepackage{latexsym}
\usepackage{fancyhdr}
\usepackage{verbatim}
\usepackage{tabularx}
\newcolumntype{Y}{>{\centering\arraybackslash}X}
\usepackage{epsfig}
\usepackage{amssymb,amsmath,amsfonts}
\usepackage{amssymb}
\usepackage{wasysym}
\usepackage{yfonts}
\usepackage{array}

\usepackage[titlenumbered,ruled, vlined]{algorithm2e}

\newcolumntype{M}[1]{>{\centering\arraybackslash}m{#1}}
\newcommand{\tikzcircle}[2][red,fill=red]{\tikz[baseline=-0.5ex]\draw[#1,radius=#2] (0,0) circle ;}%

\newcommand{\qed}{\hspace*{\fill}$\square$}

\newcommand{\be}{\begin{equation}}
	\newcommand{\ee}{\end{equation}}

\addto\captionsenglish{}
\addto\captionsenglish{}
\begin{document}
	\title{Reinforcement Learning Generation of 4-Qubits Entangled States}
	\author{Sara Giordano$^1$ and Miguel A. Martin-Delgado$^{1,2}$}
	\affiliation{$^1$Departamento de F\'{\i}sica Te\'orica, Universidad Complutense, 28040 Madrid, Spain.\\
	$^2$CCS-Center for Computational Simulation, Campus de Montegancedo UPM, 28660 Boadilla del Monte, Madrid, Spain.}
	
	\begin{abstract} 
We have devised an artificial intelligence algorithm with machine reinforcement learning (Q-learning) to construct remarkable entangled states with 4 qubits. This way, the algorithm is able to generate representative states for some of the 49 true SLOCC classes of the four-qubit entanglement states. In particular, it is possible to reach at least one true SLOCC class for each of the nine entanglement families. The quantum circuits synthesized by the algorithm may be useful for the experimental realization of these important classes of entangled states and to draw conclusions about the intrinsic properties of our universe. We introduce a graphical tool called the state-link graph (SLG) to represent the construction of the Quality matrix (Q-matrix) used by the algorithm to build a given objective state belonging to the corresponding entanglement class. This allows us to discover the necessary connections between specific entanglement features and the role of certain quantum gates, which the algorithm needs to include in the quantum gate set of actions. The quantum circuits found are optimal by construction with respect to the quantum gate-set chosen. These SLGs make the algorithm simple, intuitive and a useful resource for the automated construction of entangled states with a low number of qubits.
	\end{abstract}
	
	\maketitle

	\section{Introduction}\label{sec_I}

	The term machine learning (ML) was coined in 1959 by Arthur Samuel, an American IBMer and pioneer in the field of computer gaming and artificial intelligence \cite{Samuel_A}. ML consists essentially in the exploitation of particular algorithms of self-learning to get information from data in order to make predictions. Because of this, machine learning facilitates computers in building models from sample data in order to build decision-making processes based on data inputs \cite{Raschka, Algorithms_for_reinforcement_learning}.
	
	The applications of machine learning techniques to physics experienced a huge development in the last decades, with approaches based on topological optimization, evolutionary strategies, deep learning and reinforcement learning \cite{krenn2020computerinspired, Krenn_2016, Melnikov_2018}. In particular, the application of ML to "creative tasks", such as designing new quantum experiments, is not completely explored yet. There are a few aspects of quantum mechanics which lead researchers to think that ML and computer aided techniques offer interesting and promising perspectives in this regard \cite{Melnikov_2018}. For example, as we start to deal with many entangled particles, the Hilbert space dimension becomes so large that the problem is no more treatable without computer aid. Moreover, to build new experiments, physicists usually have to handle a huge number of variables, and ML techniques are yet in use to handle this kind of complexity \cite{krenn2020computerinspired}. Moreover, it is worth asking whether ML and artificial intelligence (AI) can boost human intuition in dealing with the intrinsic counter-intuitive nature of quantum mechanics, and if ML can also help to handle multi-dimensional and multi-partite entangled systems. An emblematic example is the Melvin algorithm developed by Krenn \textit{et al.} in 2016 \cite{Krenn_2016}. Indeed, this algorithm has uncovered solutions to previously unsolved questions and has also inspired the discovery of new scientific insights \cite{Krenn_2017}.
	
	In this work we will present an application of a ML technique to quantum entanglement, a fundamental resource for quantum computation. In particular we will employ ML to design new quantum experiments aimed at engineering quantum mechanical states with desired entanglement properties.	
	
	\section{Background}
	Machine learning algorithms can be grouped into three basic types: supervised learning, unsupervised learning and reinforcement learning (RL). Many ML reinforcement methods, such as the Projective Simulation (PS) algorithms \cite{PS}, have been extended to the quantum domain \cite{Davide} obtaining a quantum advantage for the first time over their classical counterparts. Whereas a lot of activity is being generated for quantum reinforcement learning \cite{Vedran,Lamata}, the present works remains classical with respect to the algorithm employed with the goal being the generation of quantum states. In particular, we will employ an RL algorithm called Q-learning. This latter is one of the best known RL algorithms \cite{ Algorithms_for_reinforcement_learning,Q_learn}, and, although the name may suggests otherwise, the Q does not stand for anything quantum-related, it is in fact a historical notation based on the name of the cost function to be optimized during the algorithm.
		
	\subsection{Q-learning algorithm working principles}\label{sub_I_A}
\begin{figure}
		\includegraphics[width=1\linewidth]{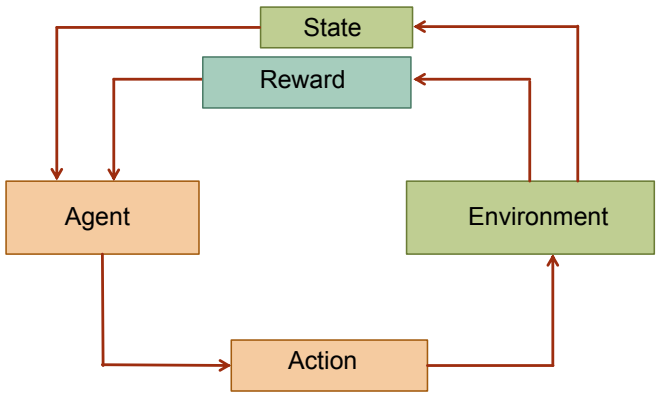}
		\caption{Flowchart of reinforcement learning working principle. It consists of an agent, an environment and a set of actions by which the agent interacts with the environment and is rewarded.}
		\label{fig1:reinforcementLearning}
	\end{figure}

\begin{figure}
		\includegraphics[width=1\linewidth]{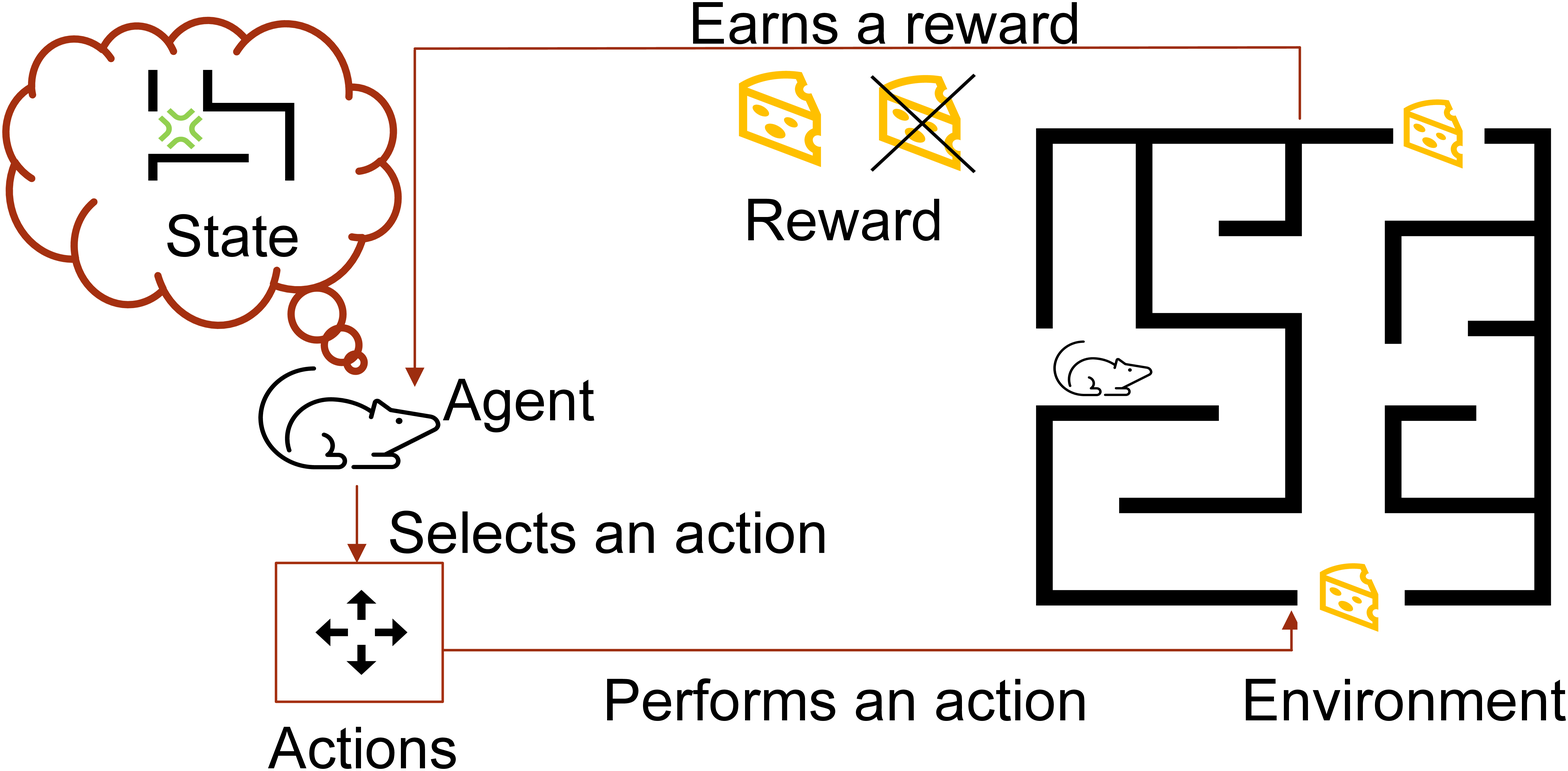}
		\caption{Q-learning scheme through the mouse-labyrinth example. In this image we represent the agent as a mouse that can move around a labyrinth, with the aim of finding the exits, where some cheese-rewards are placed. In the learning process, it selects and performs random movements starting from random positions in the labyrinth. It eventually earns a reward when it approaches the escapes and records this event. By performing random movements and by keeping track of the rewards earned, it can trace the optimal path to escape from the labyrinth, which maximizes the rewards earned.}
		\label{fig2:Mouseqlearning}
	\end{figure}

	Q-learning, like all the RL procedures, involves an agent (the algorithm itself), an environment and a set of actions by which the agent interacts with it (see Fig.~\ref{fig1:reinforcementLearning}). In the environment are placed some \textit{objectives} which the agent must reach \cite{sutton2018reinforcement}. It is a policy-free RL algorithm, i.e., there is no previous knowledge of a conduct of behavior \cite{Q_learn}. It can explore the environment by performing stochastic actions and analyzing the feedback it receives.
	By interacting with the environment, the agent changes its state and eventually earns some rewards, when finding its objectives. By keeping track of this feedback, it learns how to interact with the environment to reach the objectives while maximizing the rewards. Hence, the agent will be able to select its actions with an optimal policy \cite{Q_learn}, in the sense of the reward maximization.
	A simple illustrative description of the Q-learning algorithm is shown in Fig.~\ref{fig2:Mouseqlearning} for the mouse-labyrinth example. We can identify the following elements:
	\begin{itemize}
		\item ensemble of actions: all the possible movements that the mouse can perform,
		\item states of the agent: its positions in the labyrinth,
		\item rewards: the pieces of cheese which he can reach while walking in the labyrinth.
	\end{itemize}
	The actions of the mouse (agent), allows it to explore the labyrinth (environment). The states in which the mouse could be corresponds to all the positions in the labyrinth.	The rewards in the environment are encoded by a user defined function, which establishes the "prizes" that the agent will gain during its exploration. Usually this means that, if we want the agent to learn how to end up in specific states in the environment, we have to provide rewards every time it performs actions which lead it in those states. We will call these states the "objective states". The reward function is usually implemented as a matrix called Reward matrix (R-matrix). It has non-zero elements for \textit{state-action} pairs which lead directly to the objective states.\\
	
	While the agent explores the environment, it needs to record the rewards it earns and to recall them during the exploration. In particular it needs an instrument to keep track of the pairs \textit{state-action} which bring to rewards. In Q-learning this tool is the Quality matrix, or Q-matrix.
	The Q-learning algorithm involves two phases: training and testing \cite{Q_learn}. The training part consists in the exploration of the environment and proceeds with single \textit{episodes}. At each time $t$ an episode takes place, the agent is placed in a random state $s_t$ in the environment, from which it performs a random action $a_t$. It records the reward which it eventually obtains from the R-matrix, $R_t$. The numerical value inserted in the Q-matrix is calculated with a Bellmann equation \cite{bellman_equation}, or Q-learning formula, which allows to update the old Q-matrix value $Q(s_t,a_t)$ to the new one $Q^{new}(s_t,a_t)$. The agent assigns to each pair \textit{state-action} a quality value (Q-value), which depends not only on the reward earned in the current episode ($R_t$), but also on the rewards received in the past ones. Further details about the calculation of this Q-value will be given in the following section. Below are shown the steps consisting in a single episode:
\begin{algorithm}[h]
	\SetAlgoNoLine
	\NoCaptionOfAlgo 
	\SetKwInOut{Initialize}{Initialize}\SetKwInOut{Observe}{Observe}
	\SetKwInOut{Action}{Action}
	\Initialize{Generate a random state $s_t$}
		\Action{Select a random action $a_t$\\
		Perform the action $a_t$ from the state $s_t$}
		\Observe{Check the resulting state $s_{t+1}$}
		\If{$R_t \neq 0$}{
		Earn a reward\;}
		Update $Q(s_t,a_t)\to Q^{new}(s_t,a_t)$ with Bellman eq.\;
	\Initialize{Generate a new random state}
	\caption{Q-learning - episode of the training part}
\end{algorithm}

	At the end of the training part, the agent has updated its Q-matrix with weighted rewards associated with the pairs \textit{state-action}. Each value in the Q-matrix quantifies how much is \textit{good} to take a specific action in a certain state, in order to reach the objective with an optimal path.
	
\begin{figure}
		\includegraphics[width=1\linewidth]{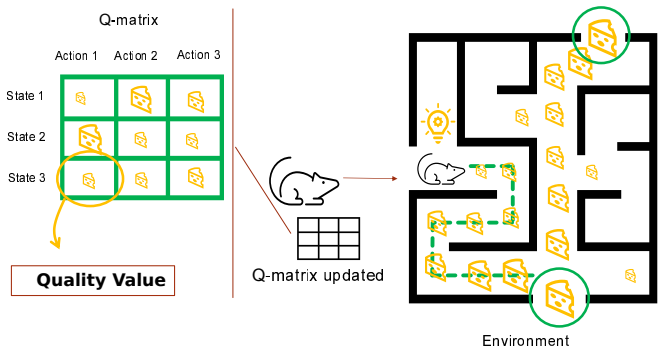}
		\caption{At the end of the training part, the mouse has updated its Q-matrix with weighted rewards associated with the pairs \textit{state-action}. This values quantify how much is \textit{good} to take a specific action in a certain state, in order to reach the objective. In the grid showed in this picture, the size of the cheese-values indicates the \textit{goodness} of an action, given the state in which it is performed. Once an initial position is established, the mouse follows his grid of \textit{state-action} pairs to select his next movement, choosing the most rewarded ones. This will carry it to the exit of the labyrinth.}
		\label{fig3:Mousetesting}
	\end{figure}

	If the agent went through a sufficiently high number of episodes, the values of its Q-matrix no longer change significantly. The criteria for this convergence must be established depending on the problem treated, but in general it is theoretically proven \cite{Q_learn,Francisco_melo} that the Q-learning algorithm converges to an optimal policy. In the testing part, by using the Q-matrix values, the agent is able to reach the \textit{a priori} established objectives by following the best rewarded pairs \textit{state-action}, starting from a chosen initial state. The agent selects from the Q-matrix the most rewarded action associated with its current state. It performs the action and changes its state, repeating this procedure until it reaches the final objective. In this way it can find the optimal path towards its goals, maximizing the earned rewards (see Fig.~\ref{fig3:Mousetesting}).\\
	
	We choose this algorithm because, even if the problem of designing quantum circuits is a deterministic problem, we need to define a code of conduct that establish which action to choose in each state, or that establish the parameters that we should evaluate in order to select an action (i.e. we are missing a policy or policy evaluation). Because of that, we chose to avoid value iteration procedures \cite{bellman2003dynamic}, or other dynamic programming methods. Moreover, we did not introduce a distance concept in our environment, and thus we are not able to assign values to the states based on their distance from the objective state.
	\subsection{Q-learning cost function}\label{sub_I_B}
	As we mentioned, the Q matrix is derived from the R-matrix, and it is updated at each episode of the training process using the Q-learning formula:
	\begin{small}
	\begin{alignat}{2}\label{eq:Q_formula}
		Q^{new}(s_{t}, a_{t})\leftarrow Q(s_{t},a_{t})(1-\alpha)+ \alpha \big(R_{t}+\gamma \: max_{\text{a}}Q(s_{t+1},a)\big).
	\end{alignat}
	\end{small}
	This expression is a Bellman equation \cite{bellman_equation}, where $Q^{new}(s_{t}, a_{t})$ expresses the quality value of a \textit{state-action} pair. In this element, the variable $t$ represent a discrete timing of the episodes.
	We can compute the value $Q^{new}(s_{t}, a_{t})$ using the prior value of the Q-matrix for that \textit{state-action} pair, $Q(s_{t}, a_{t})$, and adding to it the reward earned in the current episode, $R_{t}$. We add to this latter an evaluation of the maximum reward for the possible future actions, $max_{\text{a}}Q(s_{t+1},a)$, in fact, the $a$ index runs through all the actions that can be performed from the resulting state $s_{t+1}$. If in the past episodes the agent gained some rewards it will take into account these past rewards as a positive contribution to the $Q^{new}(s_{t}, a_{t})$ value, thanks to the quantity $max_{\text{a}}Q(s_{t+1},a)$. This means that, in order to assign a quality value to the current \textit{state-action} pair, the agent is looking at the values of the possible next steps. Indeed, with this procedure, the agent learns to follow the best rewarded path.
	These past rewards are scaled by a \textit{discount factor}, $\gamma$. 
	\begin{itemize}
		\item $\gamma$: is a real number between zero and one, $0<\gamma<1$, and sets how much the reward of future actions influences the new value $Q^{new}(s_{t}, a_{t})$.
	\end{itemize}
	If $\gamma$ is chosen close to $0$, it will prevent the algorithm to see the future rewards by making it "myopic" \cite{Q_learn}, considering only current rewards. If it is close to $1$, the past rewards will be taken much more into account than the new ones coming from the R-matrix.
	All these terms are scaled again by the so called \textit{learning rate} $\alpha$.
	\begin{itemize}
		\item $\alpha$: it is a real number between $0$ and $1$, like $\gamma$, it is set to establish the learning speed of the algorithm.
	\end{itemize}
	If it is closer to $1$ it allows the algorithm to learn quickly, as the values of the Q-matrix will be updated with a high weight, otherwise, the learning part of the algorithm is slower but is capable to "remember" better the rewards taken in the past.
	In other words, $\alpha$ defines how much we override the old values of the Q-matrix with the new ones. Hence, the setting of $\alpha$ requires to choose between two opposite strategies. The first, with $\alpha$ close to $1$, consists in a faster exploration, which easily forgets the past rewards, in favor of the new ones. The second, with $\alpha$ close to $0$, implies a slower exploration in which, however, the algorithm can remember each past reward, but it will take a longer time to explore the environment.	After setting those parameters to suitable values, the algorithm will update the Q-matrix during the training part.\\
	
	The Q-values are proven to converge to those of a $Q^*(s,a)$ function, which represents the optimal policy of the algorithm, i.e., the set of quality values that make the agent able to reach the objective with an optimal path \cite{Q_learn}. However, the training part should include ideally an infinite number of trials in order to convergence to the optimal policy \cite{Q_learn, Francisco_melo}, in practice there is a residual convergence, which is negligible. We will consider the training completed when the values of the Q-matrix remain stable under a certain threshold.
	\subsection{Entanglement basics}\label{sub_I_C}
	Entanglement is one of the most important and interesting features of quantum mechanics, it has a key role in the fields of quantum communication, quantum cryptography and entangled states are a fundamental ingredient to build quantum algorithms in quantum computation \cite{nielsen_chuang_2010, RMP02}.\\
	For pure quantum states we can say that if we have two or more systems entangled the state of each system cannot be described independently from the others. In other words, for a system with Hilbert space $H=H_A\otimes H_B$ it holds:
	\begin{gather}\label{entanglement}
		\ket{\psi}\ne \ket{\psi}_A \otimes \ket{\psi}_B,
	\end{gather}
	with $\ket{\psi}$ being an entangled state of the whole system in $H$, $\ket{\psi}$ cannot be factorized into the states of the subsystems $A$ and $B$.
	We focus our discussion on two-level quantum systems, i.e., the qubits (such as linearly polarized photons or electrons spin)), which generic state reads:
	\begin{gather}
		\ket{\phi}=\alpha\ket{0}+\beta\ket{1},
	\end{gather}
	with $\alpha$ and $\beta$ complex coefficients.	We can deal at the same time with more than one qubit, and thus we can have multi-qubit entanglement.\\
	
	The study of entanglement between qubits is of crucial importance for quantum computing and of course for the realization of modern prototypes of quantum computers \cite{nielsen_chuang_2010,RMP02}. Many studies and applications have been made to obtain desired quantum states, with particular entanglement properties, and the entanglement complexity poses major obstacles to this research field \cite{Melnikov_2018}. For this reason, many recent works employ ML techniques in order to help scientists in difficult computational tasks, but also to explore entanglement features, which are far from human intuition \cite{krenn2020computerinspired, Krenn_2016}. With such background, this work is devoted to explore the design of quantum circuits that can reproduce entangled states of qubits, with the aid of a Q-learning algorithm. In particular, we will focus our attention on entangled states of four qubits, based on the classification made previously in the literature \cite{Nielsen_conditions,Verstraete2, Verstraete_2002,slocc_in_efs, Lamata1}.
	\subsection{SLOCC classification}\label{sub_I_D}
	The complexity of quantum entangled states requires a clear picture that allows to classify them. In order to categorize different types of entanglement, we can divide the Hilbert space of a multipartite system into equivalence classes, using an operational definition of equivalence. Following the scheme in \cite{Chitambar_2014, AULBACH_2012} we can use the Local Unitary (LU) equivalence, with LU being deterministic and reversible operations. If it is possible to transform a state into another via LU operations, then the two states are LU-equivalent \cite{AULBACH_2012}. Two LU-equivalent states have the same physical properties, in particular, the same entanglement ones.
	However, the LU operations do not include all the possible operations that preserve the entanglement and can be performed experimentally. In particular they do not allow to perform joint operations on spatially separated particles. In order to classify properly the entangled states, we need to include in the conversion operations the support of classical communication \cite{Bennett_1996}. This leads to the paradigm of Local Operations assisted with Classical Communication (LOCC): quantum states are transformed by performing Local Operations (LOs) on the subsystems and allowing the transmission of classical communication between the spatially separated parties \cite{AULBACH_2012, Bennett_1996}.\\
	
	For the case of pure states, however, it has been shown \cite{Entan_monotones, LUequival} that two states are LOCC-equivalent iff they are LU-equivalent. This means that the classes defined by LOCC are the same as those defined by LU operations.
	It has been demonstrated that two pure bipartite states are LOCC-equivalent iff they have the same Schmidt coefficients \cite{Majorization, Schmidt, tenso_rank}:
	\begin{gather}
		\ket{\psi} \mathop \leftrightarrow \limits^{LU} \ket{\phi} \Leftrightarrow \ket{\psi} \mathop \leftrightarrow \limits^{LOCC} \ket{\phi} \Leftrightarrow \alpha_i = \alpha^{\prime}_i \:, \; \forall i,
	\end{gather}
	with $\ket{\psi}\to \left\{ {\alpha_i} \right\} $, $\ket{\phi}\to \left\{ \alpha^{\prime}_i\right\}$ being their Schmidt decompositions. Hence, through these LOCC we can transform one state into another deterministically (with probability of success equal to $1$), i.e the two states belong to the same LOCC entanglement class \cite{D_r_2000}, and they have the same entanglement properties. Otherwise they belong to different entanglement classes and thus have different entanglement properties. We notice that this criterion is interesting in quantum information theory because all the parties involved can use these LOCC equivalent states for exactly the same tasks \cite{D_r_2000}.\\
	
\begin{figure}
		\centering
		\includegraphics[width=0.5\linewidth]{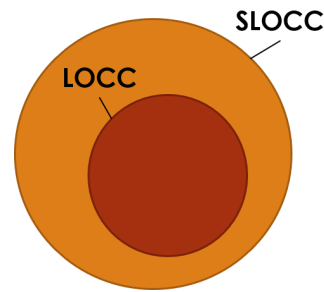}
		\caption{LOCC classification as a refinement of SLOCC classification. States which are LOCC equivalent result also SLOCC equivalent \cite{AULBACH_2012}.}
		\label{fig4:LOCC-SLOCC}
	\end{figure}

	This Schmidt criterion of classification becomes impractical when dealing with multipartite Hilbert spaces, in fact, the Schmidt decomposition is only possible for the bipartitions of a system \cite{nielsen_chuang_2010}. The most promising classification for multipartite Hilbert spaces states is the one based on the equivalence under Stochastic Local Operations and Classical Communication (SLOCC). This latter is identical to LOCC equivalence except that the interconversion of two states does not need to be deterministic, the success probability of a conversion only needs to be non-zero \cite{nielsen_chuang_2010}. Thus LOCC-equivalence implies SLOCC-equivalence, and therefore the partition of the Hilbert space into LOCC equivalence classes is a refinement of the partition into SLOCC classes, Fig.~\ref{fig4:LOCC-SLOCC}.
	The SLOCC classification is carried out mostly by means of invariants and semi-invariants \cite{AULBACH_2012,Lamata1}.	We remind that SLOCC operations cannot increase, on average, the amount of entanglement \cite{AULBACH_2012, nielsen_chuang_2010}. In particular, it is not possible to generate entangled states from separable states by SLOCC, even probabilistically \cite{AULBACH_2012, nielsen_chuang_2010}.\\

	With two qubits, there exist only two SLOCC equivalence classes, the class of separable states, and the class of entangled states. Any pure entangled state of two qubits can be converted via SLOCC into any other pure entangled state with non-zero probability. For three qubits, there exist six SLOCC classes \cite{D_r_2000}, namely, the separable class, the three bipartite entangled classes AB-C, AC-B, BC-A, the class with W-type entanglement and the class with GHZ-type entanglement \cite{D_r_2000, perfect_W}.\\
	
	For four qubits, the number of SLOCC classes seemed to become infinite \cite{D_r_2000}. This assumption was then disproved in \cite{AULBACH_2012} by D. Li \textit{et al.}, in particular they specify that it cannot be asserted that there exist infinite SLOCC classes	for four qubits. One of the reasons lies in the fact that SLOCC classes of $n$ qubits depends on at least $2\times2^n-2-8n$ real parameters \cite{Verstraete_2002}, this	lower bound allows for a finite number of SLOCC classes for $n = 4$ \cite{AULBACH_2012} in the SLOCC representation made by F. Verstraete \textit{et al.}. Meanwhile, various attempts have been made to find physically meaningful classification schemes for the four-qubit case. The technique that we will take into account is the one used in \cite{Verstraete2,Verstraete_2002}, developed also in \cite{Lamata1} and then further analyzed in \cite{slocc_in_efs}. F. Verstraete \textit{et al.} \cite{Verstraete_2002} introduced the concept of Entangled Families (EF), and nine different EFs were found, while in \cite{slocc_in_efs} 49 different SLOCC classes are identified within these EFs. This classification of four qubits states will be the basis for the application of our quantum circuit design algorithm.
	
	\subsection{SLOCC families for four qubits entangled states}\label{sub_I_E}
\begin{figure}
		\centering
		\includegraphics[width=1\linewidth]{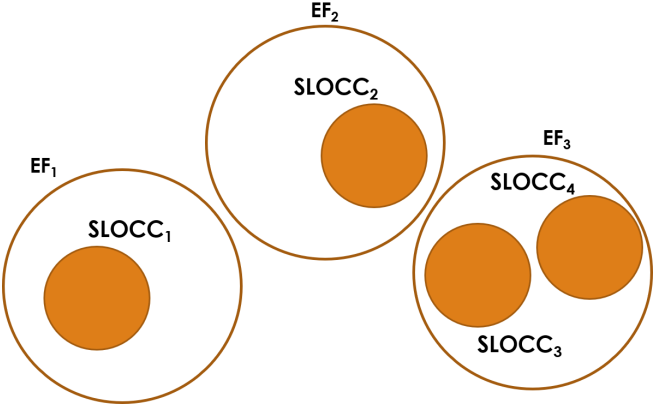}
		\caption{Graphical representation of the relationship between EFs and SLOCC classes. Even if each EF does not contain only one SLOCC class, each SLOCC class is demonstrated to belong to only one EF \cite{Verstraete_2002}.}
		\label{fig5:EFs-SLOCC}
	\end{figure}

	As we explained in the previous section, Verstraete \textit{et al.} categorize the four qubits entangled states into nine entanglement families basing the categorization on SLOCC classification \cite{Verstraete_2002}. They prove that each SLOCC equivalence class belongs to exactly one EF \cite{Verstraete_2002}. In Fig.~\ref{fig5:EFs-SLOCC} a graphical representation of this relationship is shown.
	Here we report all the nine EFs for four qubits, we adhere to the terminology used in \cite{Verstraete_2002}:
	\begin{small}
	\begin{alignat}{2}\notag
		G_{abcd}=&\frac{d+a}{2}(\ket{0000}+\ket{1111})+\frac{a-d}{2}(\ket{0011}+\ket{1100})\\\notag
		&+\frac{c-b}{2}(\ket{0110}+\ket{1001})+\frac{b+c}{2}(\ket{0101}\\ \notag
		&+\ket{1010}),\\\notag
		L_{abc_2}=&\frac{a+b}{2}(\ket{0000}+\ket{1111})+\frac{a-b}{2}(\ket{0011}+\ket{1100})\\\notag
		&+c(\ket{0101}+\ket{1010})+\ket{0110},\\\notag
		L_{a_2b_2}=&a(\ket{0000}+\ket{1111})+b(\ket{0101}+\ket{1010})+\ket{0110}\\\notag
		&+\ket{0011},\\\notag
		L_{ab_3}=&a(\ket{0000}+\ket{1111})+\frac{a+b}{2}(\ket{0101}+\ket{1010})\\\notag
		&+\frac{a-b}{2}(\ket{0110}+\ket{1001})+\frac{i}{\sqrt{2}}(\ket{0001}\\\notag
		&+\ket{0010}+\ket{0111}+\ket{1011}),\\\notag
		L_{a_4}=&a(\ket{0000}+\ket{0101}+\ket{1010}+\ket{1111})+(i\ket{0001}\\\notag
		&+\ket{0110}+i\ket{1011}),\\\notag
		L_{a_20_{3\oplus \bar{1}}}=&a(\ket{0000}+\ket{1111})+(\ket{0011}+\ket{0101}+\ket{0110}),\\\notag
		L_{0_{5\oplus\bar{3}}}=&\ket{0000}+\ket{0101}+\ket{1000}+\ket{1110},\\\notag
		L_{0_{7 \oplus\bar{1}}}=&\ket{0000}+\ket{1011}+\ket{1101}+\ket{1110},\\
		L_{0_{3\oplus\bar{1}}0_{3\oplus\bar{1}}}=&\ket{0000}+\ket{0111},
		\label{eq:nine_classes}
	\end{alignat}
	\end{small}\\
	where $a$,$b$,$c$ and $d$ are four complex parameters. The SLOCC classes which can be identified in these nine families depend on constraints applied to those four complex parameters \cite{slocc_in_efs}. The work by D. Li \textit{et al.} \cite{slocc_in_efs} distinguishes at least 49 true SLOCC entanglement classes among all the nine families. For example, for family $G_{abcd}$ they identify 13 different true SLOCC classes; for family $L_{abc_2}$, 19 true SLOCC classes and so on. They give the complete SLOCC classifications for families $L_{a_4}$, $L_{a_20_{3\oplus \bar{1}}}$, $L_{0_{5\oplus\bar{3}}}$, $L_{0_{7 \oplus\bar{1}}}$, and $L_{0_{3\oplus\bar{1}}0_{3\oplus\bar{1}}}$, but they do not grant that the other classes have been completely explored. In Tab.\ref{tab1:slocc_classes_all} we summarize the 49 true SLOCC classes, specifying the conditions on the coefficients $a$,$b$,$c$ and $d$. Here we adhere to the terminology used in \cite{slocc_in_efs}.\\
	
	Notice that, for $L_{a_20_{3\oplus \bar{1}}}$, $L_{0_{5\oplus\bar{3}}}$ and $L_{0_{7 \oplus\bar{1}}}$ in Eq.\eqref{eq:nine_classes}, there is a $1:1$ correspondence with SLOCC classes, while $L_{0_{3\oplus\bar{1}}0_{3\oplus\bar{1}}}$ does not contain four qubit entanglement, in fact it is a product state of the one-qubit state $\ket{0}$ and the three-qubit GHZ state.
	\begin{table}
		\caption{True SLOCC classes of four qubits entangled states.}\label{tab1:slocc_classes_all}
		\begin{tabular}{M{3cm} M{5.4cm}}
			\toprule
			Family, SLOCC class & Conditions on coefficients\\
			\hline
			$G_{abcd}$, A1.1        & $b=c=0, ad\neq0, a=\pm d$                      \\
			$G_{abcd}$, A1.2       & $b=c=0, ad\neq0, a\neq \pm d, a^2+d^2=0$       \\
			$G_{abcd}$, A1.3        & $b=c=0, a^2+d^2 \neq 0$                        \\
			$G_{abcd}$, A2.1        & $a=d, a \neq \pm b, b=+c, a^2+b^2=0$           \\
			$G_{abcd}$, A2.2        & $a=d, b=c, a \neq \pm b, a^2+b^2 \neq 0$       \\
			$G_{abcd}$, A3.1        & $a=d, a \neq \pm b, b=-c, a^2+b^2=0$           \\
			$G_{abcd}$, A3.2        & $a=d, b=-c, a \neq \pm b, a^2+b^2 \neq 0$      \\
			$G_{abcd}$, A4.1        & $a=d,$ either $a= \pm b$ or $\pm c, 2a^2+b^2+c^2 =0$  \\
			$G_{abcd}$, A4.2        & $a=d,$ either $a= \pm b$ or $\pm c, 2a^2+b^2+c^2  \neq 0$\\
			$G_{abcd}$, A4.3        & $a=d, a \neq \pm b, a \neq \pm c, 2a^2+b^2+c^2 = 0$\\
			$G_{abcd}$, A4.4        & $a=d, a \neq \pm b, a \neq \pm c, 2a^2+b^2+c^2 \neq 0$\\
			$G_{abcd}$, A4.5        & $a^2+b^2+c^2+d^2=0$                            \\
			$G_{abcd}$, A4.6         & $a^2+b^2+c^2+d^2 \neq 0$                       \\
			$L_{abc_2}$, B1.1        & $c=0, a=b \neq 0$                              \\
			$L_{abc_2}$, B1.2        & $c=0, a=-b \neq 0$                             \\
			$L_{abc_2}$, B1.3        & $c=0, a \neq \pm b, a^2+b^2=0$                 \\
			$L_{abc_2}$, B1.4        & $c=0, a \neq \pm b, ab \neq 0, a^2+b^2 \neq 0$ \\
			$L_{abc_2}$, B1.5        & $c=0, a \neq \pm b, ab=0$                      \\
			$L_{abc_2}$, B2.1        & $abc \neq 0, a=b, a= \pm c$                    \\
			$L_{abc_2}$, B2.2        & $abc \neq 0, a=b, a \neq \pm c, a^2+c^2=0$     \\
			$L_{abc_2}$, B2.3        & $abc \neq 0, a=b, a \neq \pm c, a^2+c^2 \neq 0$\\
			$L_{abc_2}$, B3.1        & $abc \neq 0, a=-b, a = \pm c$                  \\
			$L_{abc_2}$, B3.2        & $abc \neq 0, a=-b, a \neq \pm c, a^2+c^2=0$    \\
			$L_{abc_2}$, B3.3        & $abc \neq 0, a=-b, a \neq \pm c, a^2+c^2 \neq 0$ \\
			$L_{abc_2}$, B4.1        & $abc \neq 0, a \neq \pm b, a= \pm c, 3a^2+b^2=0$\\
			$L_{abc_2}$, B4.2        &$abc \neq 0, a \neq \pm b, a= \pm c, 3a^2+b^2 \neq 0$ \\
			$L_{abc_2}$, B4.3        & $abc \neq 0, x \neq \pm y, a^2 + b^2 + 2c^2 = 0$.\\
			$L_{abc_2}$, B4.4        & $abc \neq 0, x \neq \pm y, a^2 + b^2 + 2c^2 \neq 0$.\\
			$L_{abc_2}$, B5.1        & $c \neq 0, a=b=0$                              \\
			$L_{abc_2}$, B5.2        & $c \neq 0, a=0, b=c$                           \\
			$L_{abc_2}$, B5.3       & $c \neq 0, a=0, b \neq \pm c, b^2+2c^2=0$      \\
			$L_{a_2b_2}$, B5.4       & $c \neq 0, a = 0, b \neq \pm c, b^2 + 2c^2 \neq 0$
			\\
			$L_{a_2b_2}$, V1          & $a= \pm b \neq 0$                              \\
			$L_{a_2b_2}$, V2          & $a \neq \pm b, ab \neq 0, a^2+b^2=0$           \\
			$L_{a_2b_2}$, V3		  & $a \neq \pm b, ab \neq 0, a^2+b^2\neq 0$
			\\ 
			$L_{a_2b_2}$, V4          & $a \neq \pm b, ab=0$                           \\
			$L_{ab_3}$, R1.1        & $a=b=0$                                        \\
			$L_{ab_3}$, R1.2        & $a=b \neq 0$                                   \\
			$L_{ab_3}$, R1.3        & $a=-b \neq 0$                                  \\
			$L_{ab_3}$, R2.1        & $a=0, b \neq 0$                                \\
			$L_{ab_3}$, R2.2        & $a \neq 0, b=0$                                \\
			$L_{ab_3}$, R3.1		& $a \neq \pm b, ab \neq 0, \; 3a^2+b^2\neq 0$
			\\
			$L_{ab_3}$, R3.2        & $a \neq \pm b, ab \neq 0, 3a^2+b^2=0, b=+i\sqrt{3}$          \\
			$L_{ab_3}$, R3.2*        & $a \neq \pm b, ab \neq 0, 3a^2+b^2=0, b=-i\sqrt{3}$          \\
			$L_{a_4}$, La.1        & $a=0$                                          \\
			$L_{a_4}$, La.2        & $a \neq 0$                                     \\
			$L_{a_20_{3\oplus\bar{1}}}$& $a \neq 0$                                     \\
			$L_{0_{5\oplus\bar{3}}}$ & no conditions                                  \\
			$L_{0_{7\oplus\bar{1}}}$ & no conditions \\                                 
			\hline\hline                                               		         
		\end{tabular}
	\justifying
	In this table we show the 49 true SLOCC classes identified in \cite{slocc_in_efs} among the nine families reported in Eqs.\eqref{eq:nine_classes}, with the conditions on the parameters $a$, $b$, $c$ and $d$. Notice that the family $L_{0_{3\oplus\bar{1}}0_{3\oplus\bar{1}}}$ is not included because it is not characterized by four qubit entanglement since its representative state is a product state of the one-qubit state $\ket{0}$ and the three-qubit $\ket{GHZ}$ state \cite{slocc_in_efs}.
	\end{table}
	It is clear that, with the increasing number of qubits, the number of different SLOCC classes increase dramatically, i.e., the complexity of the entanglement grows exponentially with the number of qubits involved. To overcome this complexity and approach entangled states from a computational point of view, we decide to apply the RL algorithm described in Sec.(\ref{sub_I_A}) and Sec.(\ref{sub_I_B}), called Q-learning \cite{Q_learn}, in order to synthesize quantum circuits which can produce four qubits entangled states belonging to the above showed EFs. We will illustrate in detail in the next sections how Q-learning is exploited for our purpose.

	\section{Quantum circuits design with reinforcement learning}\label{sec_II}

	\subsection{Q-learning algorithm applied to quantum circuits design}\label{sub_II_A}
	In order to apply the Q-learning to our case, we implement the basic components of the algorithm as follows:
	\begin{itemize}
		\item the \textbf{objectives}, which we will specifically search for, are the SLOCC classes (Tab.\ref{tab1:slocc_classes_all}) included in the nine EFs [Eq.\eqref{eq:nine_classes}], hence, the representative states of the SLOCC classes of four qubits;
		\item the \textbf{environment} is located in the four qubits state space;
		\item the \textbf{actions} that the \textbf{agent} can perform consist in the application of quantum gates from a chosen set of gates;
		\item the \textbf{rewards} are encoded in a R-matrix whose entries corresponds to state-action pairs.
	\end{itemize}
	The training part consists in updating the Q-matrix whereas in the testing part the agent reaches the final objective state, producing the desired circuit. More specifically:
	\paragraph{Objectives.}
	The algorithm can only focus on one target state at a time, thus we fix a single objective state, the representative of a SLOCC class in Tab.\ref{tab1:slocc_classes_all}. In order to suit the algorithm procedure, the representative state chosen is encoded as a list of its superposition terms, e.g., if the SLOCC class is A1.1 and the representative state reads $\ket{\Psi}_{A1.1}=\frac{1}{\sqrt{2}}\left(\ket{0000}+\ket{1111}\right),$ then it is encoded as $\Psi_{A1.1}=\{0000,1111\}$. This element is a representation of the objective state that the algorithm can elaborate. In general, the state that we want to target can be written as:
	\begin{gather}\label{eq:obj_state}
		\ket{\Psi}=\sum_{j=1}^{16}\alpha_j \ket{\psi}_j,
	\end{gather}
	where $\ket{\psi}_j$ are the sixteen basis states for the four qubits state space, $\alpha_j$ are the superposition coefficients which can be either $0$ or $\neq 0$. This state, for the purpose of the algorithm, can be encoded as:
	\begin{gather}\label{eq:obj_represent}
		\Psi=\{\psi_1,\dots,\psi_n\}
	\end{gather}
	with $n$ being the number of terms of the objective quantum state $\ket{\Psi}$. Notice that in Eq.\eqref{eq:obj_represent} only the basis states $\ket{\psi_j}$ with $\alpha_j \neq 0$ are listed. We specify that there is a clear distinction between $\ket{\Psi}$, the physical representation of the quantum state, and the abstract list-object $\Psi$, which represents it for the algorithm procedure. In fact, as we will see in the next sections, after the Q-learning procedure we need to make further manipulations on the output of the algorithm to obtain the physical objective state $\ket{\Psi}$ (see Appendix).\\
	\paragraph{Environment}
	The environment of our algorithm is made up of quantum states of four qubits. We divide the Hilbert space into sub-sets characterized by states with a fixed number of terms in their superposition, i.e., single-term set, double-term set ,etc. Moreover, in order to have a finite and discrete number of states to explore, as it is done for the \textit{objectives}, we represent each state of the sub-sets as a list of its own superposition terms, without considering the states' coefficients. In fact, recording continuous coefficients would be an impossible task in terms of computational resources. However, to overcome the issues that can rise from this limitation, we will apply a post-processing procedure that allow us to tune the coefficients of the resulting state, in order to match the desired ones of the objective quantum state $\ket{\Psi}$ (see Appendix). If the representative state of the class has $n$ terms, then the algorithm will explore the sub-sets with $m$ terms, $m \leq n$. Notice that the dimension of the environment (the total dimension of the sub-sets involved) grows with the number of terms of the target state. This choice allows to explore a limited number of sub-sets.
	
	Although this method limits the possibility to explore some of the four qubits entanglement classes, our algorithm proves to be an efficient way to design quantum protocols for a significant part of them.\\
	\paragraph{Actions.}
	The actions that the agent can perform consist in the application of single quantum gates (both single qubit and multiple qubits gates). The set of gates which the algorithm can apply is established before the training part. However, depending on our necessities we can update the set of gates before each search. For example, in order to reach specific classes, we added new gates with respect to the initial ones.\\
	\paragraph{Rewards.}
	The R-matrix entries correspond to the pairs \textit{state-gate}, where the states are all the sub-sets of states with $m$ terms, which compose the environment. The only entries that carry a value $\neq 0$ are those where the \textit{gate} applied to the \textit{state} gives, as a result, the target state.
	
	\subsection{Algorithmic procedure}\label{sub_II_B}
	When a target state $\ket{\Psi}$ is established, and then encoded in $\Psi$ as a list-shaped object showed in Eq.\eqref{eq:obj_represent}, the algorithm first builds the R-matrix and initialize the Q-matrix as a zero matrix with the same shape as the R-matrix. We recall that the size of the R-matrix depends on the number of terms of the target state and on the number of gates that we decide to use.
	\paragraph{Training part.}
	During each episode of the training part, the algorithm is initialized in a random $\Psi$ state, among the ones belonging to the sub-sets involved. Then it applies to the current state a random gate, taken from the gates set. It reads the resulting state and checks if the desired target state has been reached. By means of the Q-learning cost function [Eq.\eqref{eq:Q_formula}], it updates the value of the Q-matrix entry, corresponding to the pair \textit{current state-gate}. This procedure, consisting in a single episode of the Q-learning algorithm, is repeated for a fixed number of times, established before the training part. At the end of this run, we calculate what we can call the changing rate (CR). This quantity evaluates how much the Q-matrix values change, with respect to the values before the run. We can set a threshold under which the Q-matrix is considered substantially unchanged. After a sufficiently high number of episodes, if the algorithm reaches the desired threshold, we can consider that the Q-learning is at its convergence, and thus we can proceed to the testing part. 
	
\begin{figure}
		\includegraphics[width=1\linewidth]{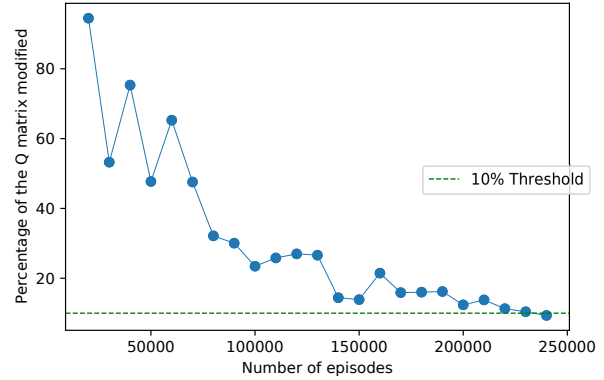}
		\caption[]{ Decreasing of the Q-matrix changing rate (CR) with the increasing number of episodes in the training part, for the search of the B1.1 SLOCC class (see Tab.\ref{tab1:slocc_classes_all}). We consider that the algorithm has successfully updated the Q-matrix when the it reaches or pass the CR threshold.}
		\label{fig6:Class_B1.1_percentage_modification}
	\end{figure}

	The number of repetitions needed to explore the whole environment depends on the dimension of this latter, which, as we mentioned, is linked to the number of terms of the searched state. In Fig.~\ref{fig6:Class_B1.1_percentage_modification} we can see an example of this convergence procedure: the threshold for the CR is set at $10\%$ and we can see how the CR of the Q-matrix decreases while number of episodes increases.
	\paragraph{Testing part.}
	In this part we check if the algorithm is able to find a suitable protocol to reach the objective state. We select an arbitrary initial state, e.g., $\ket{0000}$ encoded as $\{0000\}$. The algorithm checks the Q-matrix values in the cells that link the current state to the quantum gates, i.e., entries with coordinates ($\{0000\}$,\textit{gate}$_i)$, with $i$ scrolling all the gates of the gates-set. It then selects the action corresponding to the largest Q-value, and applies the gate to the current state. After reading the resulting state, it sets this latter as the new current state, and repeats the same procedure, until the final target state is reached. By recording the sequence of states and applied gates, it builds a quantum protocol, which starts from the arbitrarily chosen initial state and reaches to the objective one. Let us call $\ket{\Psi}_{out}$ the output state of the quantum protocol synthesized by the algorithm. The algorithm representation of the output state should be equal to the searched one $\Psi$: in other words, $\ket{\Psi}_{out}$ must have the same superposition terms showed in the $\Psi$ list, i.e., the same superposition terms of $\ket{\Psi}$. This outcome proves that the training procedure was completed successfully, and thus the algorithm is able to produce the quantum circuit that generates the target state. As we already mentioned, in order to have an output state $\ket{\Psi}_{out}$ that matches also the coefficients of the quantum state $\ket{\Psi}$ we developed a post-processing procedure showed in the Appendix.
	
	\section{Machine learning generation of entangled 4-qubit states}\label{sec_III}

	In this section we apply the Q-learning algorithm to design proper quantum protocols to achieve the four qubits entanglement classes. Using the classification in Tab.\ref{tab1:slocc_classes_all} \cite{slocc_in_efs}, we present some of the quantum circuits generated by the algorithm to reach the representative states of those classes. We find out that not all these true SLOCC classes can be handled with the aid of our algorithm. Therefore, we try to point out and explain the features of the circuits, which we manage to find for some of the classes, focusing on the specific gates that we introduced.
	
	We start our search of quantum circuits with a simple set of gates, focusing firstly on the EF $L_{0_{3\oplus\bar{1}}0_{3\oplus\bar{1}}}$, i.e., the ninth EF. Although it is characterized by only three-partite entanglement, and thus it is not a true SLOCC class, it is a good starting point to test our algorithm, due to the possibility to validate our result with pre-existing literature concerning three qubits entangled states \cite{D_r_2000, nielsen_chuang_2010}. Moreover, this family representative state has only two terms meaning that the environment we have to build is extremely limited, thus, faster to explore. Therefore, we take the representative state of the ninth family as a delightful example of the algorithm performance, and we present the results obtained with a simple set of gates, observing also its intrinsic limits. 
	
	Secondly, we show how this reduced gates-set turns out to be not successful in reaching some of the other classes. Indeed, we take as an example the seventh and eighth EFs, since they coincide with true SLOCC classes [see Eqs.\eqref{eq:nine_classes} Tab.\ref{tab1:slocc_classes_all}]. For these and other classes we are not able to produce a suitable circuit with the initial set of gates. Hence, to overcome this obstacle, we update it by adding the Toffoli gate: this allows us to reach some of the classes that were previously precluded.
	
	Afterwards, we show that also this new gate-set is not yet sufficient, as it prevents to reach some of the classes whose representatives have an odd number of terms, as it will be elucidated afterwards. The solution that we found to this problem consists in the addition of the controlled-Hadamard (C-H) gate, to our gate set.
	
	By adding new gates to our set we manage to reach a large number of SLOCC classes in Tab.\ref{tab1:slocc_classes_all}, without altering the whole structure of the algorithm.
	
	Finally, we summarize in Tab.\ref{tab3:slocc_classes_EF1_EF2}, \ref{tab4:slocc_classes_EF3_EF4_EF5}, \ref{tab5:slocc_classes_EF6_EF7_EF8} which of the 49 true SLOCC classes we manage to reach with quantum protocols built by the algorithm or with the algorithm followed by the post-processing procedure in the Appendix. Moreover, we suggest that some of the classes which we did not manage to reach, could be approached with phase-dependent gates or with new subroutines that can be queued to our procedure.
	\subsection{First set of gates}\label{sub_III_A}
	As a universal set of quantum gates we introduce the set comprising the Hadamard gate, the phase gate, the controlled-not (C-NOT) gate, and the Toffoli gate (C-C-NOT) (see Tab.\ref{tab2:universal_set}). This is equivalent to the standard universal gates set which includes Hadamard, phase gate, C-NOT and $\pi/8$ gate \cite{nielsen_chuang_2010,RMP02}.
	\begin{table}
		\caption{Universal set of quantum gates.}\label{tab2:universal_set}
		\begin{tabular}{M{3cm}  c  M{4cm}}
			\toprule
			Name & \multicolumn{1}{c}{Symbol}            & Matrix representation\\ \hline \\
			Hadamard (H)               & \Qcircuit @C=1em @R=.6em {
				& \gate{H} & \qw} & $\frac{1}{\sqrt{2}}\bigg( {\begin{array}{*{20}{c}}
					1&1\\
					1&{ - 1}
			\end{array}} \bigg)$ \\ \\
			Phase (S)                  & \Qcircuit @C=1em @R=.6em {
				& \gate{S} & \qw} & $\bigg( {\begin{array}{*{20}{c}}
					1&0\\
					0&i
			\end{array}} \bigg)$ \\ \\
			Control-not (C-NOT)         & \Qcircuit @C=1em @R=.6em {
				& \ctrl{1}  & \qw \\				
				& \targ  & \qw} & \footnotesize	$\left( {\begin{array}{*{20}{c}}
					1&0&0&0\\
					0&1&0&0\\
					0&0&0&1\\
					0&0&1&0
			\end{array}} \right)$ \normalsize \\ \\
			Toffoli (C-C-NOT)            & \Qcircuit @C=1em @R=.6em {
				& \ctrl{1}  & \qw \\
				& \ctrl{1} & \qw \\
				& \targ  & \qw} & \tiny $\quad \left( {\begin{array}{*{20}{c}}
					1&0&0&0&0&0&0&0\\
					0&1&0&0&0&0&0&0\\
					0&0&1&0&0&0&0&0\\
					0&0&0&1&0&0&0&0\\
					0&0&0&0&1&0&0&0\\
					0&0&0&0&0&1&0&0\\
					0&0&0&0&0&0&0&1\\
					0&0&0&0&0&0&1&0
			\end{array}} \right)$ \normalsize \\ \\ 
		\hline\hline
		\end{tabular}
		Universal set of quantum gates reported in \cite{nielsen_chuang_2010}.
	\end{table}
	However, as a first attempt we will use a reduced set of gates, made up of C-NOT, X gate (quantum NOT) and Hadamard. This choice is made aiming at reaching "simple classes" in the most direct and easiest way. Some of the classes, indeed, can be trivially obtained with this reduced set of gates, since approaching them with a universal set would introduce an unnecessary complication and would be more difficult in terms of computational resources. Thus, we start with a simple set and, once we meet some obstacles with this initial toolbox, we add new gates. With this approach it can be argued that we end up with a set which includes a universal set but has some redundancy, meaning that not all the gates are independent. However, by adding redundancy we gain in efficiency; indeed, a richer toolbox improves the performances of the learning part in terms of computational time.
	
	Furthermore, by including unconventional gates in the toolbox (such as controlled-Hadamard) we help the algorithm in finding optimal circuits characterized by a smaller number of gates. Moreover, handling different entanglement classes with different sets of gates, growing in complexity, allows us to better observe their entanglement properties.
	For example, some of the entanglement classes explicitly need gates which can act on three qubits (see Sec.\ref{sub_III_B} for the introduction of the Toffoli gate), while some other classes do not require this kind of gates.
	In other words, adding gates step by step helps us to point out their key role in reaching specific SLOCC classes.
	
	As anticipated, we firstly choose to analyze the ninth EF. Because its environment is extremely limited, the time needed to check if the algorithm succeeds is relatively short. In fact, since the representative of the class reads:
	\begin{gather}
		\ket{\Psi}_{L_{0_{3\otimes1}0_{3\otimes1}}}=\ket{0000}+\ket{0111},
		\label{eq:class9}
	\end{gather}
	its algorithm representation reads $\{0000,0111\}$, thus the only rewarding matrices, and quality matrices, which we need to build are those that connect \textit{actions}, i.e., application of gates, to single-term-states (ST) and double-term-states (DT). The ST and DT lists reads:
	\begin{gather*}\small
		ST=\left\{ {\begin{array}{*{20}{c}}
				{ {{\rm{0000}}} }\\
				{ {{\rm{0001}}} }\\
				{ {{\rm{0010}}} }\\
				\vdots 
		\end{array}} \right\} \qquad DT=\left\{ {\begin{array}{*{20}{c}}
				{\{ {{\rm{0000}}{\rm{,0001}}} \}}\\
				{\{ {{\rm{0000}}{\rm{,0010}}} \}}\\
				{\{ {{\rm{0010}}{\rm{,0011}}} \}}\\
				\vdots 
		\end{array}} \right\}.
	\end{gather*}
	From now on, we will refer to the representative states of the nine families as $\ket{\Psi}_J$, with $J=1,\dots,9$, and the representative states of the SLOCC classes with the notation used in Tab.\ref{tab1:slocc_classes_all} \cite{slocc_in_efs}. Thus, $\ket{\Psi}_{L_{0_{3\otimes1}0_{3\otimes1}}}=\ket{\Psi}_9$
	
\begin{figure}
		\includegraphics[width=1\linewidth]{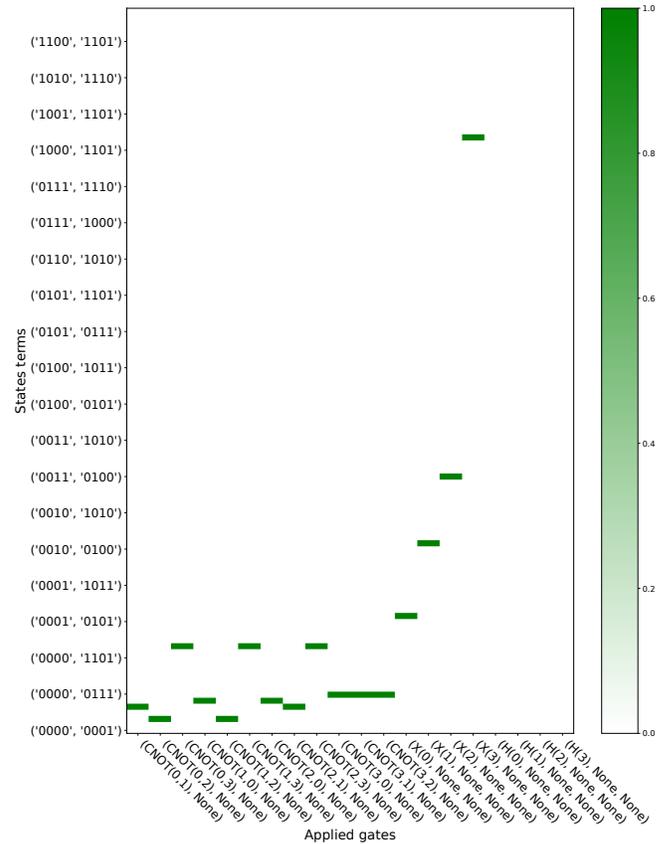}
		\caption{R-matrix for the DT states for the EF $\ket{\Psi}_9$, i.e., $L_{0_{3\oplus\bar{1}}0_{3\oplus\bar{1}}}$. The green cells represent the rewards, and they are present only for the pairs \textit{state-action} that directly link to the objective class.}
		\label{fig7:class9gridr1insight}	
	\end{figure}

\begin{figure}
		\includegraphics[width=1\linewidth]{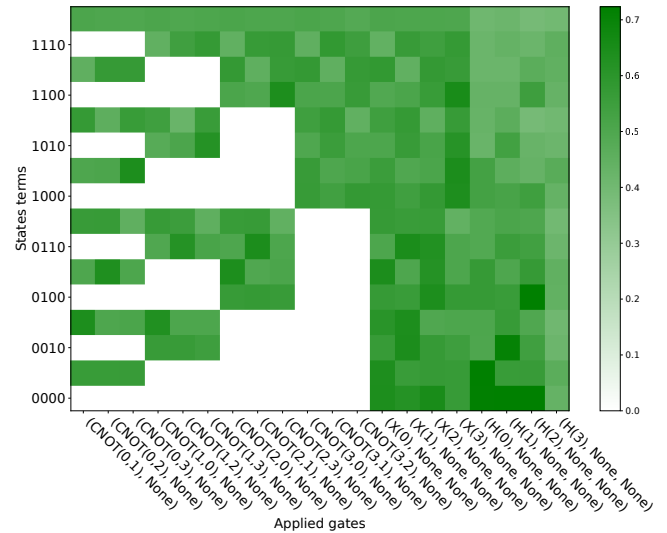}
		\caption{Q-matrix of the ninth EF, built for the ST states. The cells represent the pairs \textit{state-action} and the color-grade indicates the Q-value itself. While the R-matrix include just the single-gate rewards, this matrix contains a pattern of rewards, which help the algorithm to choose the best gate to apply when placed in a certain state.}
		\label{fig8:class9gridqinsight_ST}
	\end{figure}

	In Fig.~\ref{fig7:class9gridr1insight} we can see the R-matrix of the ninth EF of Eq.\eqref{eq:class9} for the DT states, carrying rewards only in correspondence of some \textit{state-action} pairs, highlighted in green. After training the algorithm, the Q-matrices are built: in Fig.~\ref{fig8:class9gridqinsight_ST} we can see the insight of the Q-matrix for the ST sub-set. The pattern of weighted rewards, i.e., the quality values $Q(s,a)$, is clearly visible.
	
\begin{figure}
		\centering
		\includegraphics[width=0.4\linewidth]{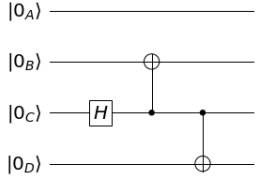}
		\caption{Quantum circuit for the $\ket{\Psi}_9$ family resulting from the reinforcement learning algorithm.}
		\label{fig9:class9circ}
	\end{figure}

	During the testing part, as explained in \ref{sub_II_B}, we choose a starting quantum state, set as the \textit{current state}, that will be the initialized state in our quantum circuit. The algorithm checks the Q-matrix values in the cells that link the current state representation to the quantum gates, i.e., if the chosen quantum state is $\ket{\psi_0}=\ket{0000}$, it checks the entries with coordinates ($\{0000\}$,\textit{gate}$_i)$, with $i$ scrolling all the gates of the gates-set. It then selects the largest Q-value in that row and applies the correspondent gate to the current state. It registers the resulting state and sets this latter as the new current state, and repeats the same procedure, until the final target state is reached.	In Fig.~\ref{fig9:class9circ} the result of the testing part is displayed in form of a quantum circuit, as the sequence of gates selected as the \textit{best valued} once. The sequential application of the gates produces the output state $\ket{\Psi}_{out}$ which, in terms of its algorithm representation $\Psi_{out}$, meets exactly the desired representative state $\ket{\Psi}_9 \to \Psi_9= \{0000,0111\}$. We will introduce in the next section an intuitive way to visualize states and gates connections.
	Based on this first successful result we may start dealing with more time-demanding classes.\\
	In Sec.\ref{sub_III_B} and \ref{sub_III_C}, we explain why we decide to add new gates to the reduced toolbox, by elucidating what kind of difficulties we encounter trying to reach some of the entanglement classes.
	\subsection{Adding the Toffoli gate}\label{sub_III_B}
	Two of the entanglement classes which the algorithm struggled to handle at this stage are the seventh and the eight:
	\begin{alignat}{2}
		\ket{\Psi}_7&=\ket{0000}+\ket{0101}+\ket{1000}+\ket{1110},\\
		\ket{\Psi}_8&=\ket{0000}+\ket{1011}+\ket{1101}+\ket{1110}.
	\end{alignat}
	For an entangled state with four terms, like $\ket{\Psi}_7$ and $\ket{\Psi}_8$, we have to explore all the sub-sets including states with $n\leq 4$ terms, i.e., ST sub-set, DT sub-set, three-terms sub-set (TT) and four-terms sub-set (FT). During the training part, even if the algorithm manages to converge in terms of the Q-matrix modifications, it shows that the rewards do not propagate outside the FT sub-set. Indeed, we noticed that, at the end of the training, only the Q-matrix related to the FT states reports weighted rewards, while the other three Q-matrices (related to TT states, DT states and ST states) remain unchanged. This means that, with the set of gates provided, there are no rewarded links between the FT sub-set and the TT, DT and ST sub-sets. 
	
	This affects our capability to reach the desired entangled state from an arbitrary quantum state in our environment. Indeed, if the agent is placed in one of the TT, DT or ST states, it will have no clue on which is the best action to take in order to reach the desired state in the four-terms sub-set, due to the fact that there are no rewards spread in that part of the environment where it is located. As a consequence, during the testing part, if the agent is placed in one of the ST, DT or TT sub-sets, it will only take random actions, because all the state-action pairs have the same quality weight (equal to zero). 
			
\begin{figure}
		\includegraphics[width=9cm]{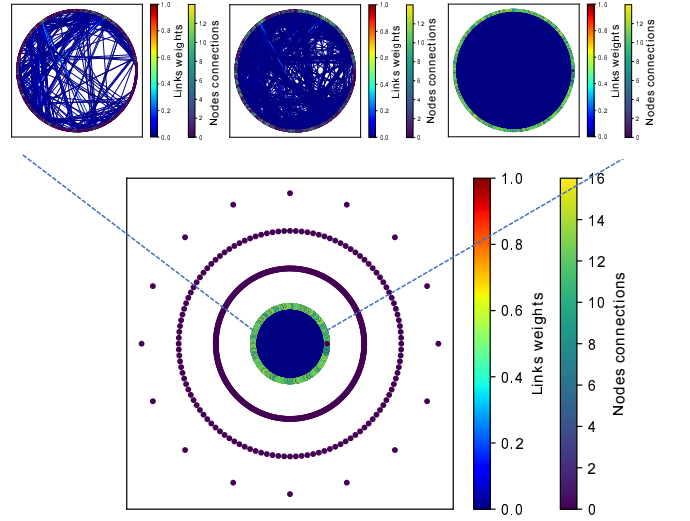}
		\caption[]{State-link graph (SLG) encoding the information of the Q-matrix for the $\ket{\Psi}_7$ SLOCC class. Each circular shell represents a sub-set of states, from the outer one to the inner one: single-terms (ST), double-terms (DT), three-terms (TT) and four-terms (FT). The nodes represent the states, while the rewarded single gate transformations are represented by the links between nodes. The color grade of the nodes refers to the number of connections that each state has with other states, belonging to the same shell or to different shells. Whereas the color of the edges indicates the weighted reward associated to that gate application. As we can see the rewards only spread inside the FT shell. On the top are showed the insights of the inner shell related to three steps (from left to right) of the progressive training procedure. We can see how the FT states are progressively connected by rewarded gates applications.}
		\label{fig10:class7graphnotoffchtotal}
	\end{figure}

\begin{figure}
		\includegraphics[width=0.5\linewidth]{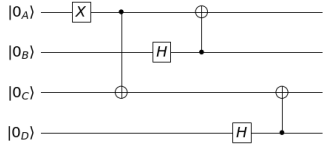}
		\caption{Quantum circuit for representative state $\ket{\Psi}_{A4.1}(a=0)$}
		\label{fig11:classA4.1circ}
	\end{figure}

\begin{figure}
		\centering
		\includegraphics[width=0.5\linewidth]{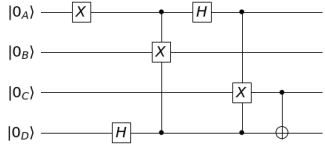}
		\caption{Quantum circuit built for $\ket{\Psi}_7$ with the aid of the Toffoli gate.}
		\label{fig12:class7circ}
	\end{figure}

\begin{figure}
		\includegraphics[width=9.5cm]{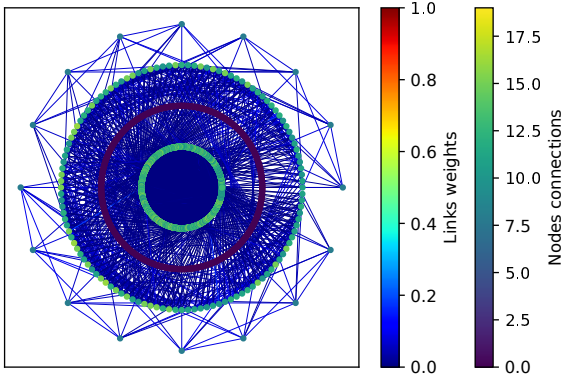}
		\caption{State-link graph (SLG) for the objective state $\ket{\Psi}_{7}$ after the introduction of the Toffoli gate. In this graph the four shells are completely connected by weighted rewards, as opposed to the situation in Fig.~\ref{fig10:class7graphnotoffchtotal}.}
		\label{fig13:class7graphtotal}
	\end{figure}

	We can visualize the connections between quantum states, encoded in the Q-matrix, with the state-link graph (SLG) shown in Fig.~\ref{fig10:class7graphnotoffchtotal}. The nodes of the graph represent the environment states and the links which connect two states correspond to the rewarded application of a single gate. Different concentric shells correspond to different sub-sets, i.e., to sub-sets which states have different number of terms. The innermost shell is the one related to the four-terms states: it is the shell where the representative state of the class is located. The other shells, from the outer one to the inner one, are the ST states, the DT states and the TT states. The color grade of the nodes refers to the number of connections that each state has with other states and the colors of the links correspond to the Q-values. We can see that the nodes of the outer shells have no rewarded connections, while the four-terms shell has links between its states.
	This behavior indicates that the algorithm could not spread the reward pattern in the Q-matrix outside a certain region of the environment. If we want our algorithm to find the optimal path starting from an arbitrary initial state, we have to provide shortcuts that can connect different shells with a rewarded link.
	Notice that this obstacle cannot be traced back to the number of terms in the superposition of the objective state, because for other classes with four terms the algorithm worked efficiently. As an example we can look at the entanglement family $G_{abcd}$, in particular, the SLOCC class $A4.1$ in Tab.\ref{tab1:slocc_classes_all}, where the parameters are set as: $a=d,\; a= \pm b, \;2a^2+b^2+c^2 =0$, with $a=0$:
	\begin{gather}\notag
		\ket{\Psi}_{A4.1}(a=0)=\ket{0101}+\ket{1010}+\ket{0110}+\ket{1001}\\ \notag
		\downarrow \text{algorithm representation}\\
		\Psi_{A4.1}(a=0)=\{0101,1010,0110,1001\}
	\end{gather}
	For this class the rewards spread over all the states sub-sets and the algorithm is able to find the circuit showed in Fig.~\ref{fig11:classA4.1circ}, starting from the state $\ket{\psi_0}=\ket{0000}$.
	The reason for this behavior lies on the particular terms of the superposition characterizing the two classes $\ket{\Psi}_8$ and $\ket{\Psi}_7$: we can see that both have one or more terms with three qubits in the state $\ket{1}$. In fact $\ket{\Psi}_8$ is in a superposition of $\ket{1011}$, $\ket{1101}$ and $\ket{1110}$, and $\ket{\Psi}_7$ includes $\ket{1110}$. To account for this feature it becomes necessary to introduce the Toffoli gate (C-C-NOT). It performs a quantum not on a target qubit only if the two control qubits are in the state $\ket{1}$ at the same time. Indeed, acting on three qubits the Toffoli gate generates this type of superposition without creating or canceling other terms in the state. 
	For our purpose, the Toffoli gate allows the algorithm to reach superpositions between base elements which include terms with three qubits in the state $\ket{1}$. Indeed, after the introduction of C-C-NOT in the set of gates, the algorithm manages to spread the reward and finds paths that allow to reach $\ket{\Psi}_7$ starting from an arbitrary quantum state. In Fig.~\ref{fig12:class7circ} we can see the quantum circuit for the seventh EF, with $\ket{\psi_0}=\ket{0000}$.
	We can also observe the Q-matrices pattern of rewards, once the Toffoli is introduced: in fact, the state-link graph (SLG) appears completely filled with rewarded links that connect different shells Fig.~\ref{fig13:class7graphtotal}.
	
	With the addition of the Toffoli gate we almost reached a universal set of gates, which comprehends: Hadamard gate, Z gate, C-NOT, and Toffoli gate \cite{nielsen_chuang_2010}. However, due to the fact that we are not keeping track of the amplitudes, we do not need to add the phase gate Z.
	\subsection{Adding the Controlled-Hadamard gate}\label{sub_III_C}
	Proceeding with the analysis of the SLOCC classes we find that, at this stage, the algorithm is unable to reach some states with an odd number of terms in their superposition. In particular, states which representative has three terms, such as the representatives of classes $B1.1$,$B1.2$ and $B5.1$.
	Again we decide to add a gate to our toolbox, in order to reach these SLOCC classes. The best candidate that can help us is the Hadamard gate, because it introduces superposition. Indeed, its action reads:
	\begin{gather}
		H\ket{0}\rightarrow\frac{1}{\sqrt{2}}\left(\ket{0}+\ket{1}\right),\quad
		H\ket{1}\rightarrow\frac{1}{\sqrt{2}}\left(\ket{0}-\ket{1}\right),
	\end{gather}
	In terms of a quantum optics device, by taking the polarization as our degree of freedom and considering linearly polarized photons, the Hadamard corresponds to a polarization rotator which performs a rotation of $\pi/4$, i.e., a half-wave plate. Indeed, horizontal and vertical polarization can be encoded, respectively, as $\ket{0}$ and $\ket{1}$ of the computational basis. A qubit subject to the action of a Hadamard gate is analogous to a photon passing through a half-wave plate: it is projected onto the basis $\{\ket{+}=1/\sqrt{2}\left(\ket{0}+\ket{1}\right), \ket{-}=1/\sqrt{2}\left(\ket{0}-\ket{1}\right)\}$ \cite{scully_zubairy_1997}.
	
	Applying Hadamard to a single-term state we obtain a double-terms state. Of course, applying it to a double-term state we may obtain a three-terms state. However, when we look at what happens when a Hadamard is applied to a double-terms state, \textit{in an homogeneous superposition}, we can see that there are just a few possible outcomes.
	
	We have two different cases, depending on the entanglement. If the state is entangled and it is in an homogeneous superposition, the action of the Hadamard gate can only lead to one type of result, e.g., on one of the Bell's states it reads:
	\begin{gather}
		H(A)(\ket{00}+\ket{11})\to \ket{00}+\ket{10}+\ket{01}-\ket{11}.
	\end{gather}
	where $H(A)$ is the Hadamard acting on the first qubit. 
	A similar superposition of four terms can be obtained from the action of one Hadamard gate on the other three Bell states. The other possibility is that the two-qubit state, on which the Hadamard acts is not entangled, then the result can be summarized in the following way:
	\begin{itemize}
		\item[\textit{i}] Action on the factorizable qubit\\ \begin{alignat}{2}\notag
			H(A)(\ket{01}+\ket{00})\to& H(A)\ket{0}_A(\ket{1}_B+\ket{0}_B)\\ \notag
			\to& \ket{01}+\ket{11}+\ket{00}+\ket{01}.
		\end{alignat}
		\item[\textit{ii}] Action on the non-factorizable qubit\\ \begin{alignat}{2}\notag
			H(B)(\ket{01}+\ket{00})\to& \ket{0}_AH(B)(\ket{1}_B+\ket{0}_B)\\ \notag
			\to& \ket{00}.
		\end{alignat}
	\end{itemize}
	This means that the Hadamard gate, acting on the two qubits states in homogeneous superpositions, can only double the number of terms (from two to four terms) or halves it, returning to a single-term state.
	This discussion can be generalized to the three and four qubit case. Indeed, as the Hadamard gate is a single-qubit gate, if we want to obtain a recombination of terms, i.e., to sum two or more terms, we need superpositions of terms, which differ from each other for a single qubit and the Hadamard gate should act on that specific qubit:
	\begin{gather}
		H(A)(\ket{01}+\ket{11})\to \ket{01}+\ket{11}+\ket{01}-\ket{11}\to \ket{01}.
	\end{gather}
	If we apply the Hadamard gate to a two-terms state of four or three qubits with terms which differ for more than one qubit, we will obtain always four terms as a result, because the sum of the terms would not be possible, keeping in mind that the Hadamard acts on only one qubit at a time. This observation leads to the following results regarding the four qubits states, analogues to those listed above for the two-qubit states:
	\begin{itemize}
		\item[\textit{i*}] The two terms differ from each other for more than one qubit, and the Hadamard gate acts on one of them
		\begin{alignat}{2}\notag
			H(B)(\ket{0100}-\ket{0010})\to& \ket{0000}-\ket{0100}\\ \notag
			&-\ket{0010}+\ket{0110}.
		\end{alignat}
		
		We obtain a four-terms superposition from a two-terms one.
		\item[\textit{ii*}] The two terms differ one another for more than one qubit and the Hadamard gate acts on the factorizable ones (analogue to the case \textit{i})
		
		\begin{alignat}{2}\notag
			H(A)(\ket{0100}+\ket{0010})\to& \ket{0100}+\ket{1100}\\ \notag
			&+\ket{0010}+\ket{1010}.
		\end{alignat}
		
		We obtain a four-terms superposition from a two-terms one.
		\item[\textit{iii*}] The two terms differ for a single qubit and the Hadamard gate acts on the factorized ones (this case is not different from the previous one \textit{ii*}, due to the fact that the Hadamard gate can only change one qubit at a time)
		\begin{flalign}\notag
			H(A)(\ket{0100}+\ket{0000})\to & 
			\ket{0100}+\ket{1100}\\ \notag
			&+\ket{0000}+\ket{1000}.
		\end{flalign}
		
		We obtain a four-terms superposition from a two-terms one.
	\end{itemize}
	If the two terms differ by only one qubit, which corresponds to the qubit targeted by the Hadamard gate, then the result can be trivially referred to the two-qubit case \textit{ii}, without loss of generality:
	\begin{itemize}
		\item[\textit{iv*}] The two terms differ for a single qubit and the Hadamard gate acts on it
		\begin{flalign}\notag
		H(D)(\ket{0000}+\ket{0001})\to & \ket{0000}+\ket{0001}\\ \notag 
		&+\ket{0000}-\ket{0001}\\ \notag 
		\to & \ket{0000}.
		\end{flalign}
		
		We obtain a state which is no more in a superposition, a single-term state from a two-terms one.
	\end{itemize}
	In polarization terms, we are performing a rotation from the diagonal basis $\ket{+}_D=1/\sqrt{2}\left(\ket{0}_D+\ket{1}_D\right)$, $\ket{-}_D=1/\sqrt{2}\left(\ket{0}_D-\ket{1}_D\right)$ to the basis $\ket{0}_D,\ket{1}_D$. Thus, the Hadamard gate \textit{cannot provide three terms in a superposition}.
	
\begin{figure}
		\includegraphics[width=9.5cm]{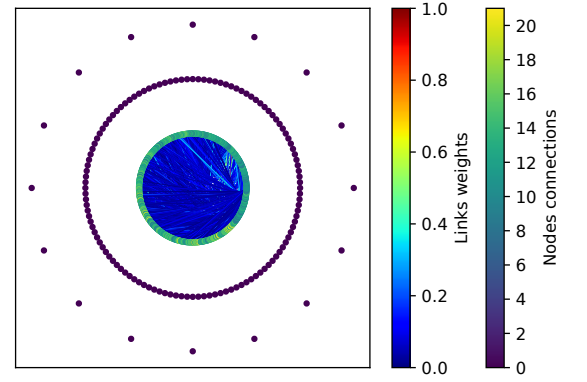}
		\caption{The state-link graph (SLG) for the objective state $\ket{\Psi}_{B1.1}$ without the C-H gate added, showing that the shells related to ST states (outer shell) and DT (next-to-outer shell) states are not connected by rewarded single gates applications.}
		\label{fig14:classB1.1graphtotalnotoffCH}
	\end{figure}

\begin{figure}
		\includegraphics[width=9.5cm]{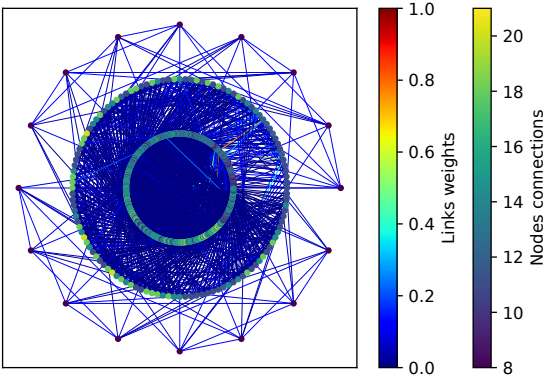}
		\caption{The state-link graph (SLG) for the objective state $\ket{\Psi}_{B1.1}$ with the C-H gate added showing how the shells, corresponding to all number of terms, link with one another.}
		\label{fig15:class_B1.1_graphtotal}
	\end{figure}

\begin{figure}
		\includegraphics[width=0.4\linewidth]{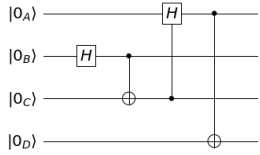}
		\caption{Quantum circuit obtained for the SLOCC class B1.1 (Tab.\ref{tab1:slocc_classes_all}) generated with the reinforcement learning algorithm.}
		\label{fig16:classB1.1circ}
	\end{figure}

\begin{figure}
		\includegraphics[width=9.5cm]{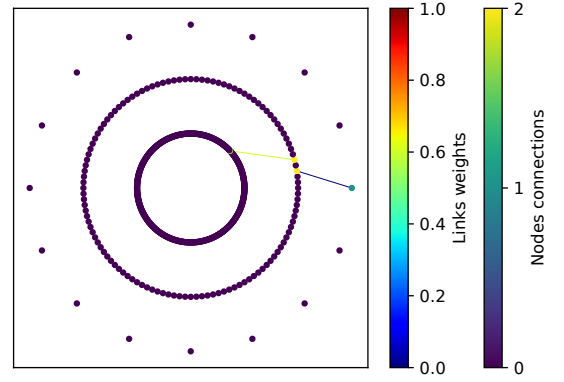}
		\caption{Quantum circuit for the class B1.1 in the SLG representation. These links represent connections created by the single gates of the protocol in Fig.~\ref{fig16:classB1.1circ}. The starting state on the right is $\Psi_0=\{0000\}$.}
		\label{fig17:classB1.1path}
	\end{figure}

	The discussion becomes different if we consider states of four terms and we want to reach five terms. In that case the Hadamard gate, together with the action of NOT or C-NOT gates, can recombine terms and a state with five terms can be built from a four-terms state.
	Since some of the entanglement classes that we want to reach have representative states with three terms, we choose to add a new gate to the previous set of gates, namely, the Controlled-Hadamard (C-H) gate. To see the effectiveness of this addition, we take as an example the SLOCC class B1.1, whose representative state is $\ket{\Psi}_{B1.1}=\ket{0000}+\ket{1111}+\ket{0110}$ \cite{slocc_in_efs}. In Fig.~\ref{fig14:classB1.1graphtotalnotoffCH} we can see the SLG related to this entanglement class without the addition of the C-H gate.
	We notice that the TT states are isolated from the others, as it happened before with $\ket{\Psi}_7$. With the C-H included in the toolbox we can overcome this obstacle and we are able to generate states with three terms in their superposition. Without loss of generality, we can make an example of the action of a C-H on a simple two-qubits state, omitting normalization coefficients:
	\begin{gather}
		C\text{-}H(B,A)(\ket{01}_{AB}+\ket{10}_{AB})\rightarrow \ket{01}+\ket{11}+\ket{10}.
	\end{gather}
	In this example the C-H gate has qubit B as the control qubit and qubit A as te target. Due to the non-symmetric action of this gate we can reach states with an odd number of terms with a shortcut which connects even-terms shells and odd-terms shells in the SLG. In Fig.~\ref{fig15:class_B1.1_graphtotal} we can see how the graph related to $\ket{\Psi}_{B1.1}$ is now connected by edges that allow to reach the objective state.
	We report the quantum circuit to generate $\ket{\Psi}_{B1.1}$ in Fig.~\ref{fig16:classB1.1circ}, and the corresponding optimal path in a graph form in Fig.~\ref{fig17:classB1.1path}.

	\subsection{Complete and incomplete exploration of the entanglement classes}\label{sub_III_E}
	We summarize in Tabs.\ref{tab3:slocc_classes_EF1_EF2}, \ref{tab4:slocc_classes_EF3_EF4_EF5}, \ref{tab5:slocc_classes_EF6_EF7_EF8} the sub-group of the 49 classes, showed in Tab.\ref{tab1:slocc_classes_all}, which we managed to approach with our reinforcement learning algorithm. We recall that these SLOCC classes are identified from the original nine entanglement families in Eqs.\eqref{eq:nine_classes}, with constraints on the complex parameters $a$, $b$, $c$ and $d$ \cite{slocc_in_efs}. The colored dots in the \textit{Feasibility} column have the following meaning:
	\begin{itemize}
		\item[ ]\tikzcircle[fill=black!40!green]{4pt} $\rightarrow$ class that we are able to reach with the set of gates \{X, H, C-NOT, C-C-NOT, C-H\}, without the post-processing procedure,
		\item[ ] \tikzcircle[fill=blue]{4pt} $\rightarrow$ class that can be reached with the previously mentioned set of gates and the post-processing procedure described in the Appendix, 
		\item[ ] \tikzcircle[fill=yellow]{4pt} $\rightarrow$ class that requires post-processing procedure shown in the Appendix and phase gates addition.
	\end{itemize}
	For the SLOCC classes marked with a dark green dot (\tikzcircle[fill=black!40!green]{4pt}), we are able to produce the quantum circuits generating their representative state exploiting only the Q-learning procedure. We showed some of them in the previous sections.
	These classes have representative states that are homogeneous superpositions with real coefficients, i.e., their terms have all the same real coefficients. For these classes, we managed to find suitable protocols, reported in Sect.~\ref{Quantum_Circuits}, which can reproduce their representative states. We consider these results optimal given the settings of the algorithm.
	
	The classes with a blue mark (\tikzcircle[fill=blue]{4pt}) required the post-processing procedure, described in the Appendix. Indeed, even though the circuit produced by the Q-learning procedure has an output state with the same terms of the representative one, we need to tune the coefficients of this output state, to match those of the desired state. 
	
	The classes with a yellow mark (\tikzcircle[fill=yellow]{4pt}) have representative states in which one or more terms have imaginary coefficients. Hence, our algorithm is able to reach states with the right terms of the superposition but our post processing procedure is not able to match the coefficients in the end, as it is designed only to tune real coefficients. These states can be reached with a proper addition of the phase gates (S), the control-phase gates (C-S) or the $\pi/8$ gates. By means of them we can rearrange the phases of the coefficients accordingly, in order to achieve the desired representative state. In any case, our algorithm can be used as a fruitful starting point.
	
	There are some classes in Tab.\ref{tab1:slocc_classes_all} that are not included in this summary. This is because they are not fully classified \cite{slocc_in_efs}, meaning that they could either be true SLOCC classes or not. Let us recall that this feasibility categorization is based on the capabilities of our algorithm and it is intrinsically linked to its working principles. Thus, among the classes listed in this section, we consider \textit{completely explored} the classes marked in dark green and blue, while we consider the others as partially explored, which could be finalized along the lines mentioned here or by further developments.
	
	We divide the SLOCC classes into three tables. The first one, Tab.\ref{tab3:slocc_classes_EF1_EF2}, refers to the SLOCC classes that derive from the first two EFs in Eqs.\eqref{eq:nine_classes}. We notice that the four-qubit GHZ state is included in the first EF, the representative state of class A1.1. being,
	\begin{gather}
		\ket{\text{GHZ}}_{4}=\frac{1}{\sqrt{2}}\left(\ket{0000}+\ket{1111}\right).
	\end{gather}
		\renewcommand{\arraystretch}{1.6}
	
	\begin{table}[h]
		\caption{Feasibility categorization.}
		\begin{tabularx}{\linewidth}{M{2cm}  M{4.5cm}  M{2.1cm}}
			\toprule
			Family, SLOCC class & Conditions on parameters & Feasibility \\ \hline
			$G_{abcd}$, A1.1        & $b=c=0, ad\neq0, a=\pm d$                      & \tikzcircle[fill=black!40!green]{4pt}\\
			$G_{abcd}$, A1.2        & $b=c=0, ad\neq0, a\neq \pm d, a^2+d^2=0$       & \tikzcircle[fill=yellow]{4pt}\\
			$G_{abcd}$, A2.1        & $a=d, a \neq \pm b, b=+c, a^2+b^2=0$           & \tikzcircle[fill=yellow]{4pt}\\
			$G_{abcd}$, A3.1        & $a=d, a \neq \pm b, b=-c, a^2+b^2=0$           & \tikzcircle[fill=yellow]{4pt}\\
			$L_{abc_2}$, B1.1        & $c=0, a=b \neq 0$                              & \tikzcircle[fill=blue]{4pt}\\
			$L_{abc_2}$ , B1.2       & $c=0, a=-b \neq 0$ & \tikzcircle[fill=blue]{4pt}\\
			$L_{abc_2}$, B1.3        & $c=0, a \neq \pm b, a^2+b^2=0$                 & \tikzcircle[fill=yellow]{4pt}\\
			$L_{abc_2}$, B1.5        & $c=0, a \neq \pm b, ab=0$                      & \tikzcircle[fill=blue]{4pt}\\
			$L_{abc_2}$, B2.1        & $abc \neq 0, a=b, a= \pm c$                    & \tikzcircle[fill=blue]{4pt}\\
			$L_{abc_2}$, B2.2        & $abc \neq 0, a=b, a \neq \pm c, a^2+c^2=0$     & \tikzcircle[fill=yellow]{4pt}\\
			$L_{abc_2}$, B3.1        & $abc \neq 0, a=-b, a = \pm c$                  & \tikzcircle[fill=blue]{4pt}\\
			$L_{abc_2}$, B3.2        & $abc \neq 0, a=-b, a \neq \pm c, a^2+c^2=0$    & \tikzcircle[fill=yellow]{4pt}\\
			$L_{abc_2}$, B5.1        & $c \neq 0, a=b=0$                              & \tikzcircle[fill=blue]{4pt}\\
			$L_{abc_2}$, B5.2        & $c \neq 0, a=0,b=c=1$  & \tikzcircle[fill=blue]{4pt}\\
			$L_{abc_2}$, B5.3        & $c \neq 0, a=0, b \neq \pm c, b^2+2c^2=0$      & \tikzcircle[fill=yellow]{4pt}\\
			\hline\hline   
		\end{tabularx}
		\justifying
		In this table we report the feasibility categorization of our algorithm and post-processing procedure for the SLOCC classes belonging to the EFs $G_{abcd}$ and $L_{abc_2}$.
		\label{tab3:slocc_classes_EF1_EF2}
	\end{table}
	In Tab.\ref{tab4:slocc_classes_EF3_EF4_EF5} the SLOCC classes that derive from the third, fourth and fifth families in Eqs.\eqref{eq:nine_classes} are reported.
		\renewcommand{\arraystretch}{1.6}
	\begin{table}[h]
		\caption{Feasibility categorization.}
		\begin{tabularx}{\linewidth}{M{2cm}  M{4.5cm}  M{2.1cm}}
			\toprule
			Family, SLOCC class & Conditions on parameters & Feasibility \\ \hline
			$L_{a_2b_2}$, V1          & $a= \pm b \neq 0$                              & \tikzcircle[fill=blue]{4pt}\\
			$L_{a_2b_2}$, V2          & $a \neq \pm b, ab \neq 0, a^2+b^2=0$           & \tikzcircle[fill=yellow]{4pt}\\
			$L_{a_2b_2}$, V4        & $a \neq \pm b, ab=0$                           & \tikzcircle[fill=blue]{4pt}\\
			$L_{ab_3}$, R1.1        & $a=b=0$                                        & \tikzcircle[fill=black!40!green]{4pt}\\
			$L_{ab_3}$, R1.2        & $a=b \neq 0$                                   & \tikzcircle[fill=yellow]{4pt}\\
			$L_{ab_3}$, R1.3        & $a=-b \neq 0$                                  & \tikzcircle[fill=yellow]{4pt}\\
			$L_{ab_3}$, R2.1        & $a=0, b \neq 0$                                & \tikzcircle[fill=yellow]{4pt}\\
			$L_{ab_3}$, R2.2        & $a \neq 0, b=0$                                & \tikzcircle[fill=yellow]{4pt}\\
			$L_{ab_3}$ , R3.2        & $a \neq \pm b, ab \neq 0, 3a^2+b^2=0$          & \tikzcircle[fill=yellow]{4pt}\\
			$L_{ab_3}$ , R3.2*        & $a \neq \pm b, ab \neq 0, 3a^2+b^2=0$          & \tikzcircle[fill=yellow]{4pt}\\
			$L_{a_4}$, La.1        & $a=0$                                          & \tikzcircle[fill=yellow]{4pt}\\
			$L_{a_4}$, La.2        & $a \neq 0$                                     & \tikzcircle[fill=yellow]{4pt}\\
			\hline\hline
		\end{tabularx}
		\justifying
		Here we show the feasibility for the EFs $L_{a_2b_2}$, $L_{ab_3}$ and $L_{a_4}$. We can see that, due to the complex coefficients of the EF $L_{ab_3}$ in Eqs.\eqref{eq:nine_classes}, the SLOCC classes related to this EF are not approachable with our post-processing procedure.
		\label{tab4:slocc_classes_EF3_EF4_EF5}
	\end{table}

	In Tab.\ref{tab5:slocc_classes_EF6_EF7_EF8} are reported the SLOCC classes belonging to the remaining families in Eqs.\eqref{eq:nine_classes}, the sixth, the seventh and the eighth families, since the ninth family does not contain four qubits entanglement. 
	\begin{table}[h]
		\caption{Feasibility categorization.}
		\begin{tabularx}{\linewidth}{M{2cm}  M{4.5cm}  M{2.1cm}}
			\toprule
			Family, SLOCC class & Conditions on parameters & Feasibility  \\ \hline
			$L_{a_20_{3\oplus\bar{1}}}$& $a \neq 0$& \tikzcircle[fill=blue]{4pt}\\
			$L_{0_{5\oplus\bar{3}}}$& no constraints&\tikzcircle[fill=black!40!green]{4pt}\\
			$L_{0_{7\oplus\bar{1}}}$& no constraints& \tikzcircle[fill=blue]{4pt}\\
			\hline\hline         
		\end{tabularx}
		\justifying
		The last three SLOCC classes that we report are among those showed as examples in this paper, $\ket{\Psi}_6$,$\ket{\Psi}_7$ and $\ket{\Psi}_8$.
		\label{tab5:slocc_classes_EF6_EF7_EF8}
	\end{table}

\subsection{Quantum circuits for SLOCC classes of 4 qubits}\label{sec_IV} \label{Quantum_Circuits}

	This subsection summarizes the main results of this work. We report in Tabs.\ref{tab6:quantum_protocols_1} and \ref{tab7:quantum_protocols_2} the quantum circuits that we find with our reinforcement learning algorithm for some of the classes listed in Tabs.\ref{tab3:slocc_classes_EF1_EF2}, \ref{tab4:slocc_classes_EF3_EF4_EF5}, \ref{tab5:slocc_classes_EF6_EF7_EF8}, marked with a blue (\tikzcircle[fill=blue]{4pt}) and a dark green dot (\tikzcircle[fill=black!40!green]{4pt}).
	We can see that, in some cases the algorithm uses the Toffoli gate and the C-H gate extensively: this is due to the optimization of the circuital length.
		\begin{table}[h]
			\caption{Quantum protocols.}
			\begin{tabular}{M{3.3cm} M{5cm}}
				\toprule
				\begin{tabular}{@{}M{3cm}@{}}SLOCC class\\ Representative state\end{tabular}          & Quantum protocol \\ \hline
				\begin{tabular}{@{}M{3cm}@{}}A1.1 \\ $\ket{0000}+\ket{1111}$\end{tabular}	&	\vspace{0.1cm}		\begin{minipage}{0.30\textwidth}
					\includegraphics[width=0.6\linewidth]{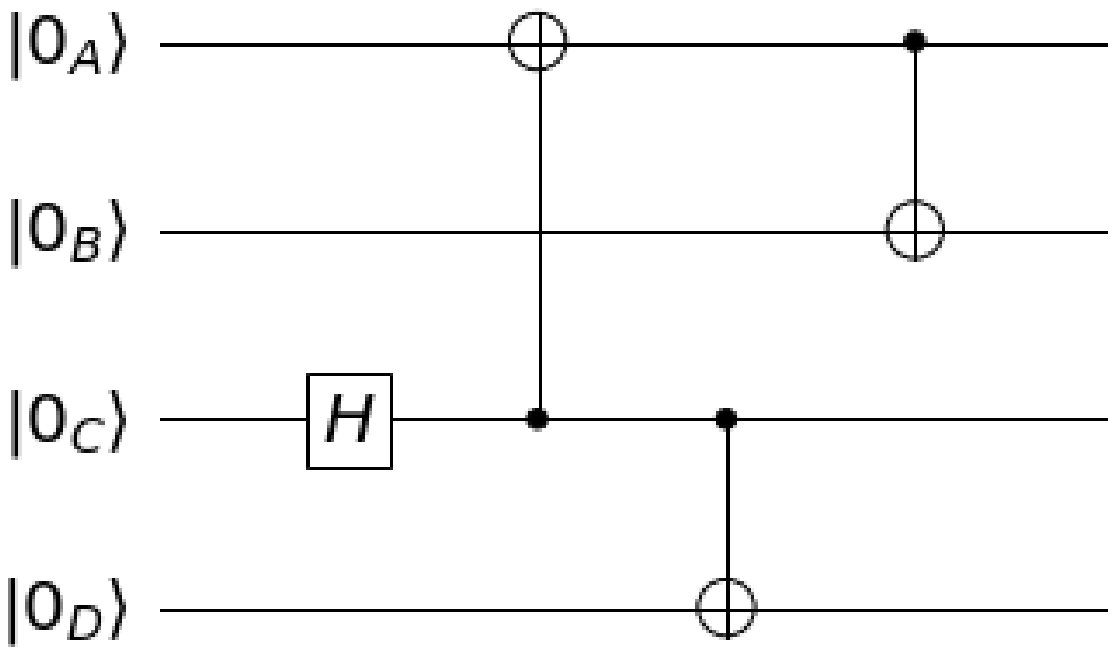}\label{fig18:Class_A1.1_circ}
				\end{minipage}
				\\ \hline
				\begin{tabular}{@{}M{3cm}@{}}B1.1\\$\ket{0000}+\ket{1111}+\ket{0110}$\end{tabular}	& \vspace{0.2cm}
				\begin{minipage}{.30\textwidth}
					\includegraphics[width=0.6\linewidth]{Class_B1.1_circ.eps}
				\end{minipage}
				\\ \hline
				\begin{tabular}{@{}M{3cm}@{}}B1.2\\$\ket{0011}+\ket{1100}+\ket{0110}$\end{tabular}	& \vspace{0.2cm}
				\begin{minipage}{.30\textwidth}
					\includegraphics[width=0.8\linewidth]{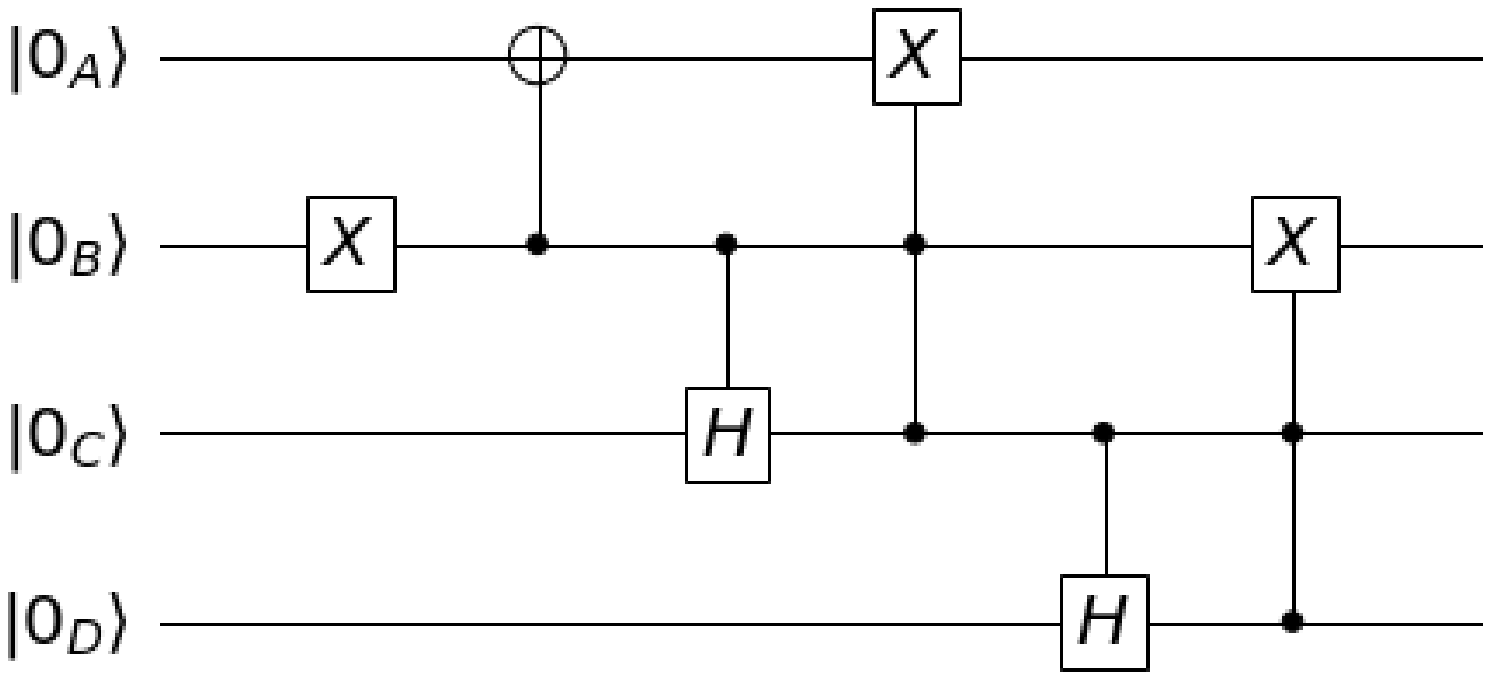}\label{fig19:Class_B1.2_circ}
				\end{minipage}
				\\ \hline
				\begin{tabular}{@{}M{3cm}@{}}B2.1\\$\ket{0000}+\ket{1111}+\ket{0101}+\ket{1010}+\ket{0110}$\end{tabular}	& \vspace{0.2cm}
				\begin{minipage}{.30\textwidth}
					\includegraphics[width=0.8\linewidth]{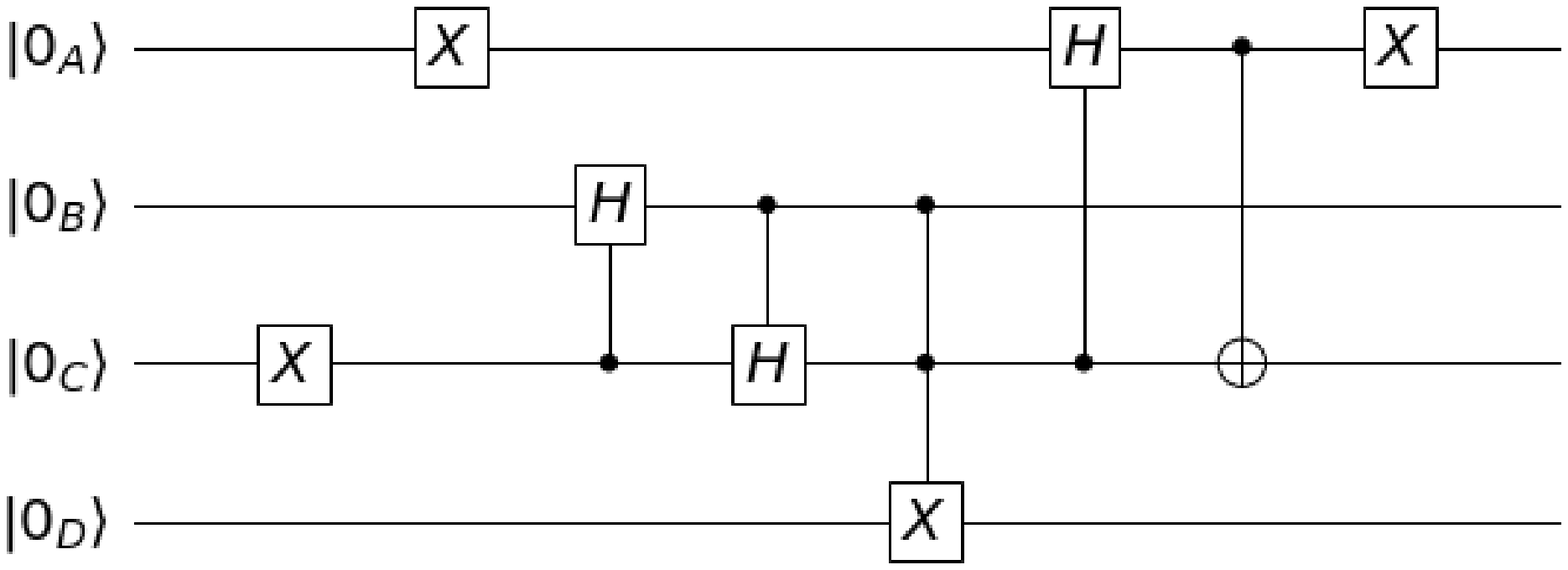}\label{fig20:Class_B2.1_circ}
				\end{minipage}
				\\ \hline
				\begin{tabular}{@{}M{3cm}@{}}B3.1\\$\ket{0011}+\ket{1100}+\ket{0101}+\ket{1010}+\ket{0110}$\end{tabular}	& \vspace{0.2cm}
				\begin{minipage}{.30\textwidth}
					\includegraphics[width=0.8\linewidth]{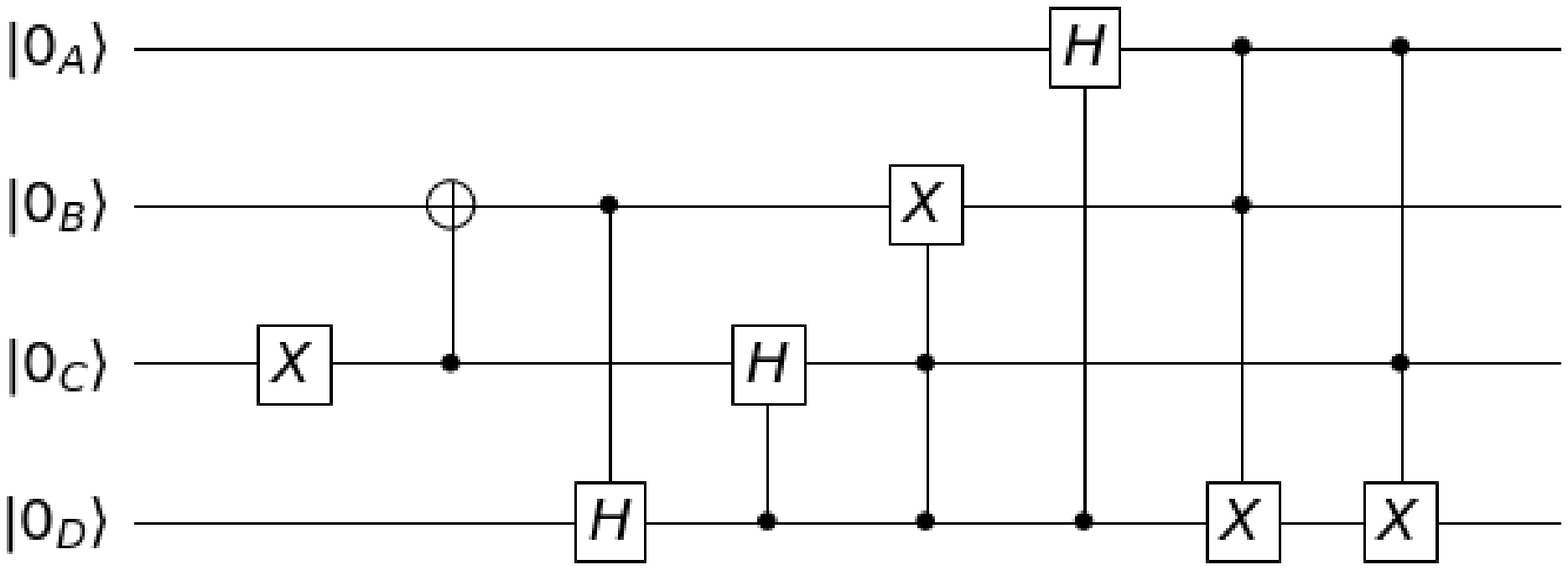}\label{fig21:Class_B3.1_circ}
				\end{minipage}
				\\ \hline\hline
			\end{tabular}
			\justifying
			Quantum protocols designed with the algorithm for some of the SLOCC classes of four-qubit entangled states.
		\label{tab6:quantum_protocols_1}
		\end{table}
		\begin{table}[h]
			\caption{Quantum protocols.}
			\begin{tabular}{>{\centering}M{3.3cm} M{5cm}}
				\toprule
				\begin{tabular}{@{}M{3cm}@{}}SLOCC class\\ Representative state\end{tabular}          & Quantum protocol \\ \hline
				\begin{tabular}{@{}M{3cm}@{}}B5.1\\$\ket{0101}+\ket{1010}+\ket{0110}$\end{tabular}	& \vspace{0.2cm}
				\begin{minipage}{.30\textwidth}
					\includegraphics[width=0.8\linewidth]{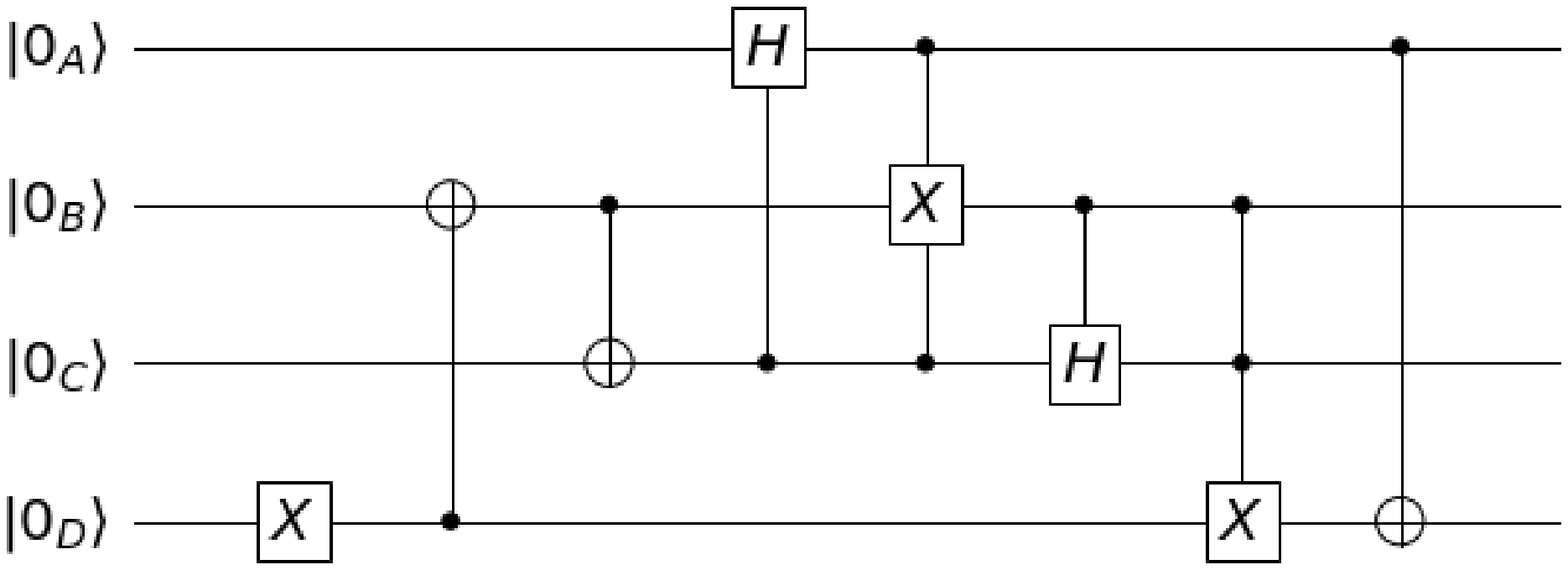}\label{fig22:Class_B5.1_circ}
				\end{minipage}
				\\ \hline
				\begin{tabular}{@{}M{3cm}@{}}V4\\
				$\ket{0000}+\ket{1111}+\ket{0110}+\ket{0011}$\end{tabular}	& \vspace{0.2cm}
				\begin{minipage}{.30\textwidth}
					\includegraphics[width=0.8\linewidth]{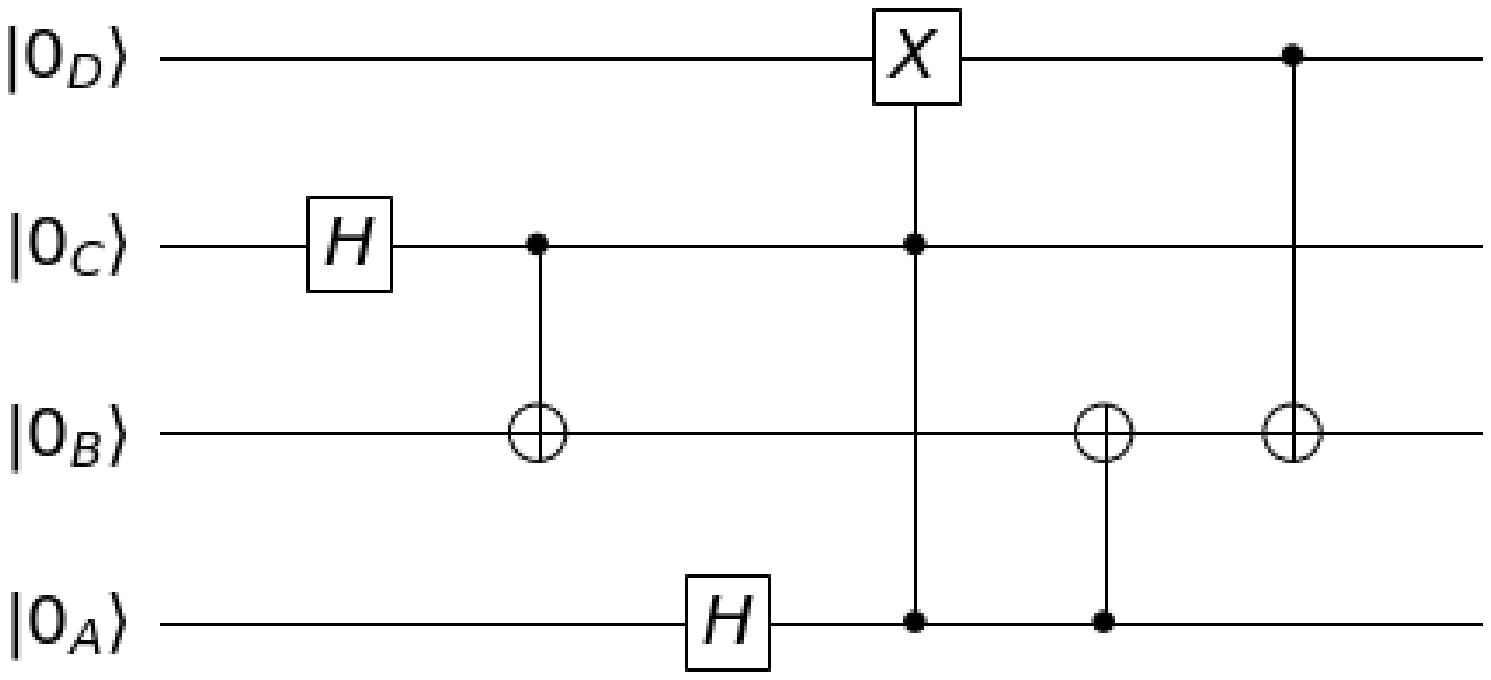}\label{fig23:Class_V4_circ}
				\end{minipage}
				\\ \hline
				\begin{tabular}{@{}M{3cm}@{}}R1.1\\
				$\ket{0001}+\ket{0010}+\ket{0111}+\ket{1011}$\end{tabular}	& \vspace{0.2cm}
				\begin{minipage}{.30\textwidth}
					\includegraphics[width=0.8\linewidth]{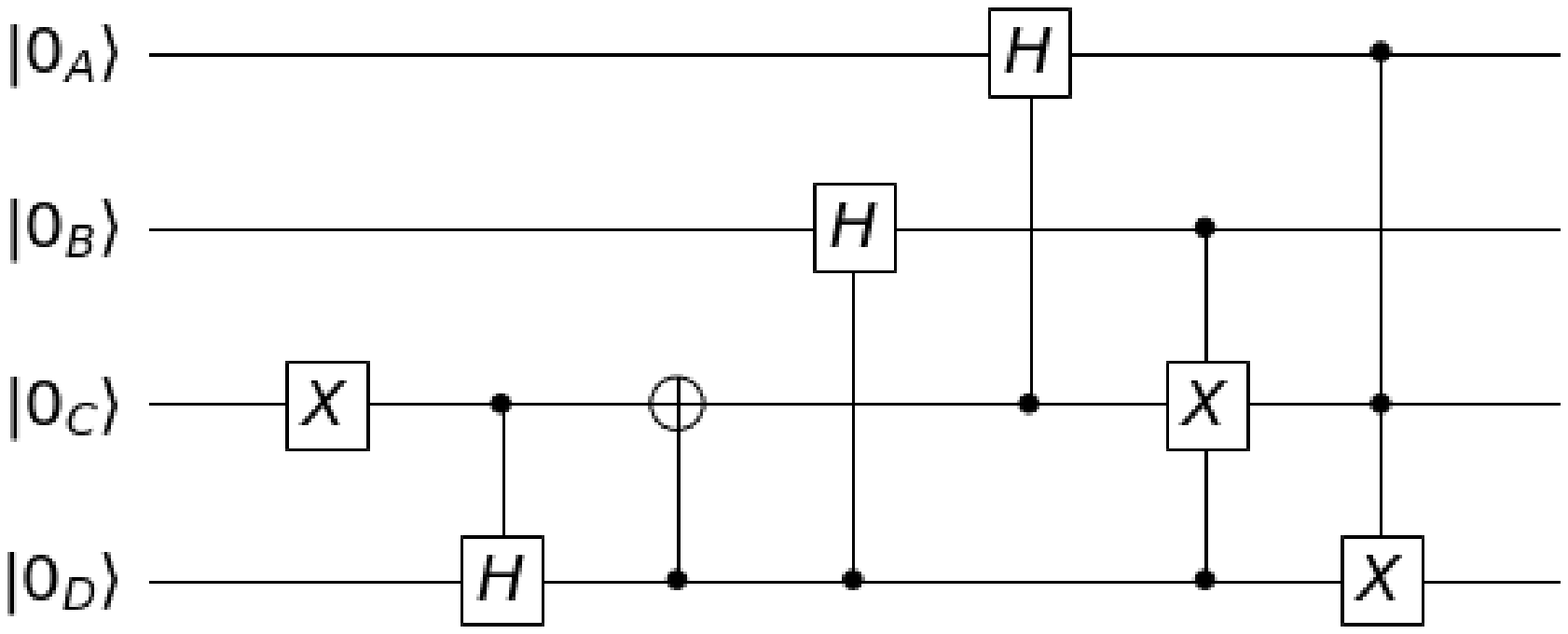}\label{fig24:Class_R1.1_circ}
				\end{minipage}
				\\ \hline 
				\begin{tabular}{@{}M{3cm}@{}}La.1\\
				$\ket{0001}+\ket{0110}+\ket{1011}$\end{tabular}	& \vspace{0.2cm}
				\begin{minipage}{.30\textwidth}
					\includegraphics[width=0.8\linewidth]{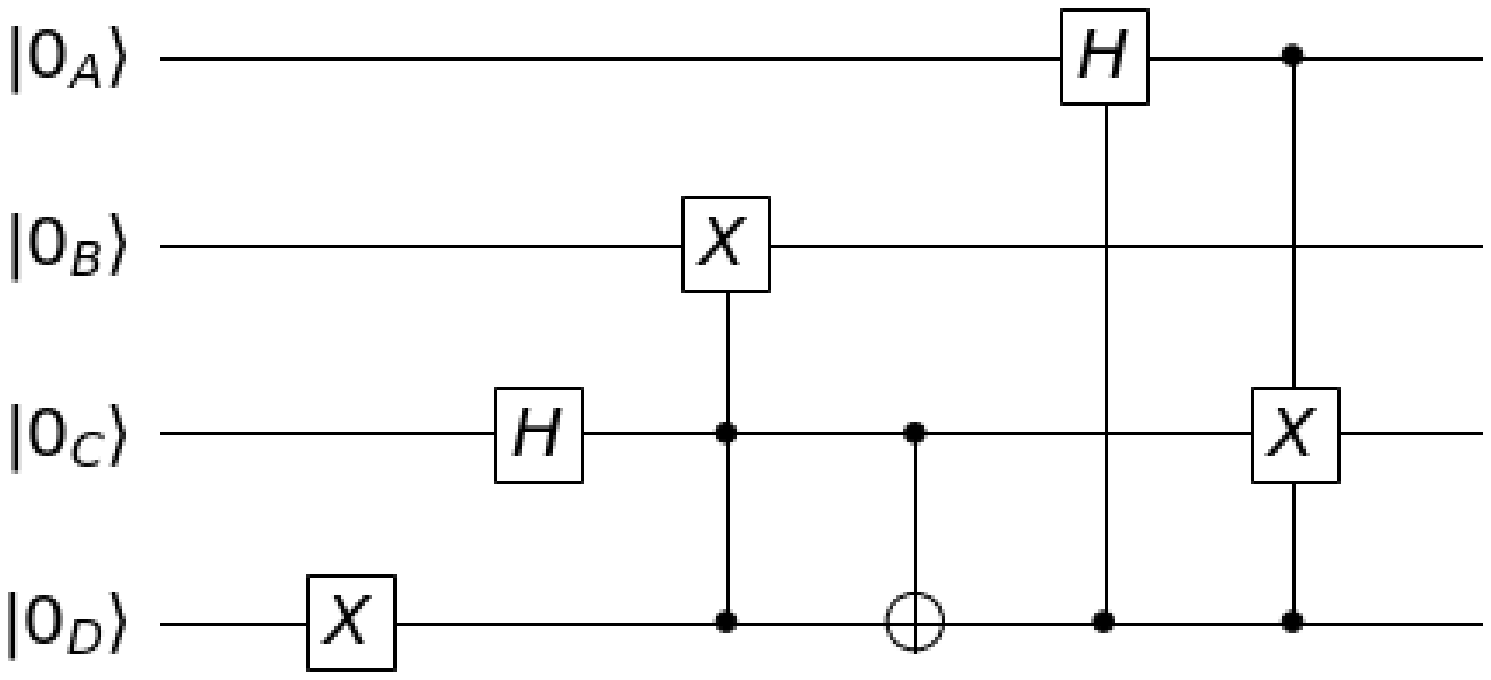}\label{fig25:Class_La.1_circ}
				\end{minipage}
				\\ \hline
				\begin{tabular}{@{}M{3cm}@{}}$L_{a_20_{3\oplus\bar{1}}}$\\
				$\ket{0000}+\ket{1111}+\ket{0011}+\ket{0101}+\ket{0110}$\end{tabular}	& 
				\vspace{0.2cm}
				\begin{minipage}{.30\textwidth}
					\includegraphics[width=0.8\linewidth]{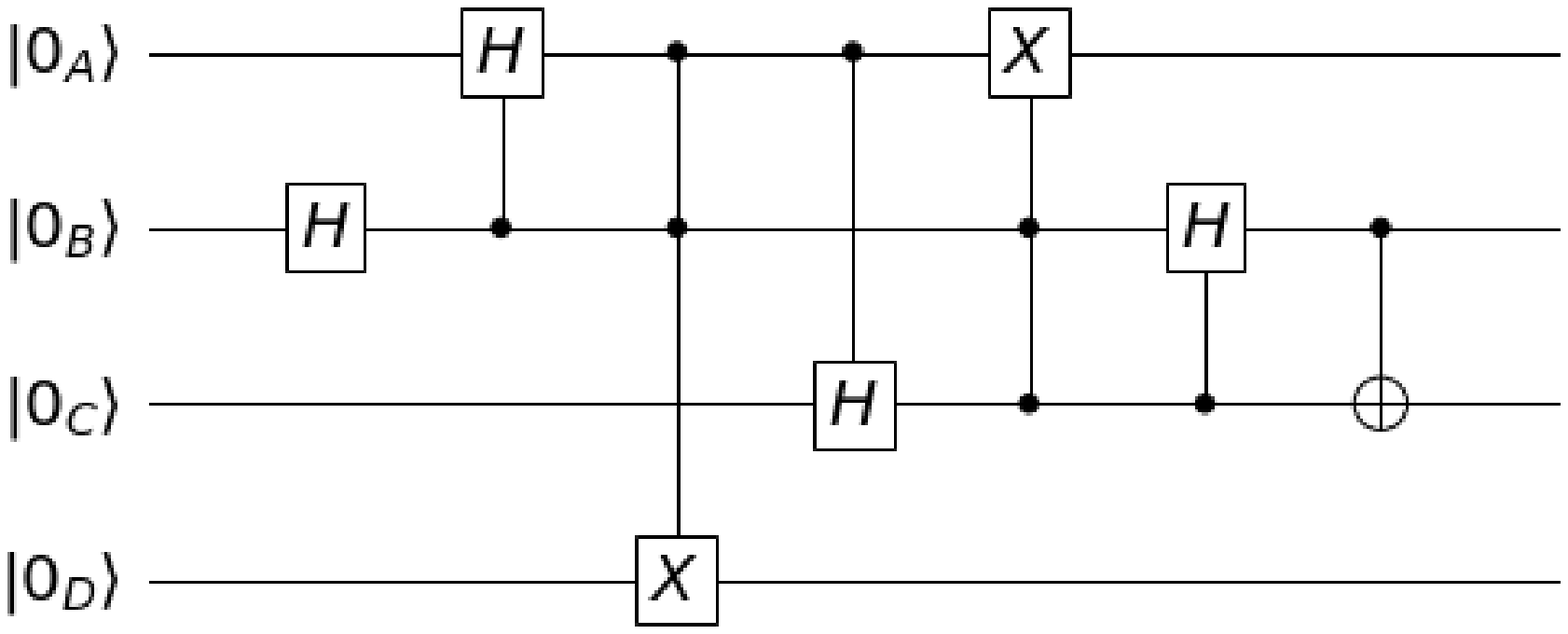}\label{fig26:Class_6_circ}
				\end{minipage}
				\\ \hline
				\begin{tabular}{@{}M{3cm}@{}}$L_{0_{5\oplus\bar{3}}}$\\
				$\ket{0000}+\ket{0101}+\ket{1000}+\ket{1110}$\end{tabular}	& 
				\vspace{0.2cm}
				\begin{minipage}{.30\textwidth}
					\includegraphics[width=0.8\linewidth]{Class_7_circ_NO_CH.eps}
				\end{minipage}
				\\ \hline
				\begin{tabular}{@{}M{3cm}@{}}$L_{0_{7\oplus\bar{1}}}$\\
				$\ket{0000}+\ket{1011}+\ket{1101}+\ket{1110}$\end{tabular}	& 
				\vspace{0.2cm}
				\begin{minipage}{.30\textwidth}
					\includegraphics[width=0.8\linewidth]{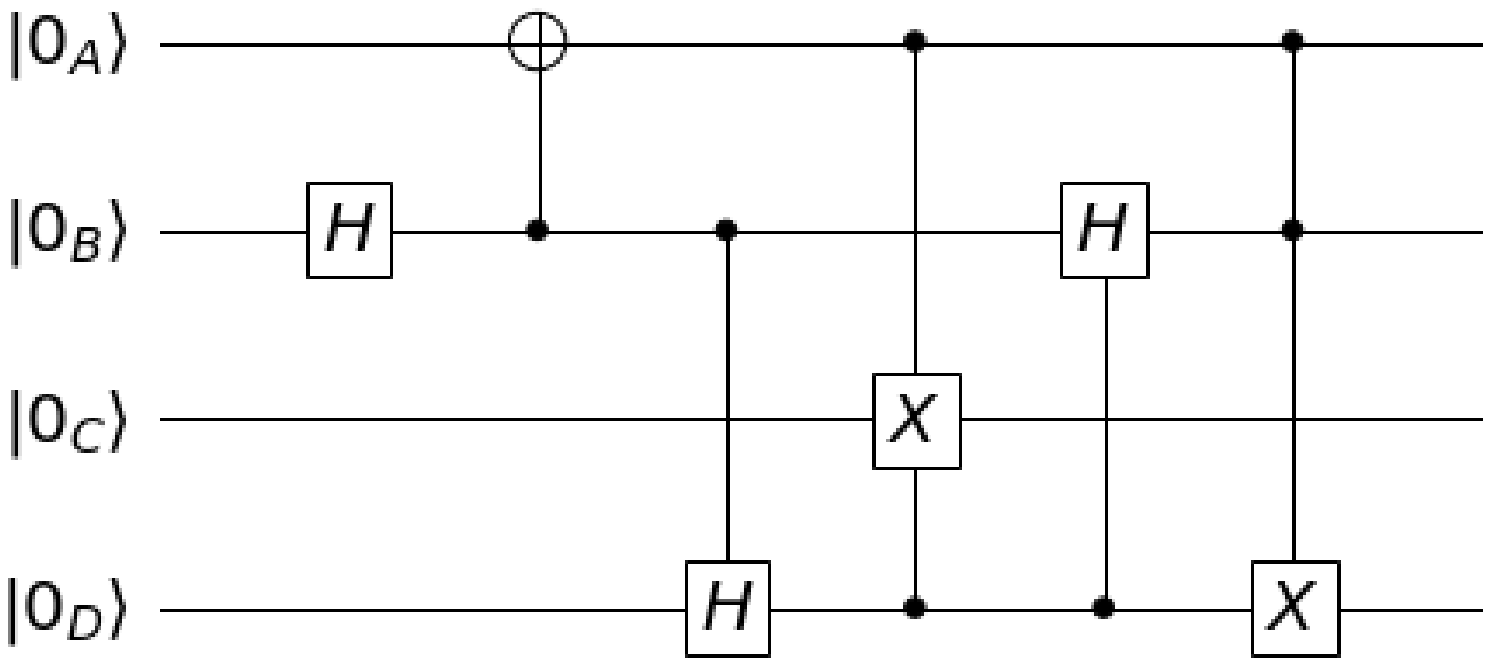}\label{fig27:Class_8_circ}
				\end{minipage}
				\\ \hline\hline
			\end{tabular}
			\justifying
			Quantum protocols for some of the SLOCC classes of four-qubit entangled states.
			\label{tab7:quantum_protocols_2}
		\end{table}
	\section{Conclusions}\label{sec_V}
	We exploited the potentiality of the off-policy Q-learning algorithm that provided us a tool for state preparation in quantum computing. We focused on the interplay between the chosen sets of gates, the encoding method and visualization in SLGs, and the entanglement SLOCC classes. The idea to use a single gate reward carried to highlight the role of some specific single gates in reaching states with some qualitative entanglement features. We pursued the idea of giving an operative indication on the complexity of qubits state preparation, in terms of circuit composition.
	
	We have shown that with our implementation of the Q-learning algorithm, we manage to successfully build quantum protocols able to generate representative states for some of the 49 true SLOCC classes of the four-qubit entanglement states. In particular, we are able to reach at least one true SLOCC class for each of the nine entanglement families. Further, we observe that many of the other SLOCC classes can be approached by adding other quantum gates to the set, and modifying accordingly the algorithm in order to use them. Therefore, this machine learning algorithm is useful in reaching a large number of four-qubit entangled states, and could be employed to better understand their properties and to devise new procedures to construct them in a real experiment. Moreover, our method in principle allows to deal with $n$ qubit states, $4$-party quantum states with multilevel systems, i.e., we could in future works extend it from qubits to qutrits or higher, thereby it can in principle produce further new results in unknown territory.
	
	Furthermore, we can discover new connections between specific entanglement features and the role of certain quantum gates. In this sense, thanks to its simplicity and intuitiveness, the Q-learning algorithm turns out to be widely profitable for these kind of tasks. Something similar to this was discovered with the Melvin algorithm \cite{Krenn_2016,Krenn_2017} for a low number multidimensional (a few qutrits, etc.,) quantum states. It is conceivable that those multidimensional quantum states can be addressed with the tools developed in this work.
	
	Due to our limited computational resources and the intricacies of some of the entangled states addressed, we have not completed the full generation of all the 49 classes of entangled 4 qubit states. Thus, a possible next step to take to reach the representatives of the remaining classes would be to make the algorithm capable of handling states in a non-homogeneous superposition, with complex coefficients, without the aid of a post processing method. In this way, the number of SLOCC classes for which we could be able to provide protocols will further increase. Therefore, even if the Q-learning in its current form is not suitable for the very complete exploration of the nine entanglement families, it provides reasonable clues as to how to attempt to handle the unsolved cases, if further capabilities are included in the resource toolbox of the reinforcement Q-learning algorithm.
	
	We have devised a graphical tool called the state-link graph (SLG) to represent the construction of the Q-matrix for a given objective state belonging to one of the entanglement classes. See examples in figures \ref{fig10:class7graphnotoffchtotal}, \ref{fig13:class7graphtotal} etc. These graphs are very useful to detect whether the learning algorithm is exploring the set of multiple terms needed to reconstruct the objective state. This way, when it is detected that some of the shells in SLG are not connected, it is an indication that our gate-set chosen is not rich enough so as to build the given state. Then, it means that it is the moment to enlarge the quantum gate-set. This is precisely the process that we have followed to synthesize the quantum circuits found in Tabs.\ref{tab6:quantum_protocols_1}, \ref{tab7:quantum_protocols_2}.
	
	Some of the results obtained for the synthesis of 4-qubit states with remarkable entanglement properties, such as those in Tabs.\ref{tab6:quantum_protocols_1}, \ref{tab7:quantum_protocols_2}, may be useful to investigate statements about the local and realistic properties of our universe with experimental means as was originally proposed with the Melvin algorithm \cite{Krenn_2016,Krenn_2017}. In our Q-learning algorithm, we do not need to know about the concept of Schmidt coefficients as the Melvin algorithm does. The reinforcement machine learning algorithm does not rely on previous knowledge nor on often flawed intuition.

	By construction, the quantum circuits found with our reinforcement learning algorithm are optimal with respect to the quantum gate-set chosen. This is guaranteed by the convergence of the training part and the subsequent construction of the quantum circuits in the testing part relying on optimal state-action pairs found for the Q-matrix. This result is useful for the automated quantum circuit synthesis (QCS) where optimal implementations of quantum algorithms are designed from quantum logic gates belonging to known universal sets \cite{QCS1,QCS2,QCS3,QCS4,QCS5}. These are the type of automated tasks needed to construct quantum compilers. Machine learning methods to synthesize optimal circuits for continuous variable quantum computation have been proposed with photonics architectures \cite{ nichols2019designing, Arrazola}.

	It is worth noticing that our reinforcement learning algorithm is not scalable as the number of entangled qubits increases since the number of multiple terms needed to construct the objective states and the environment, grows exponentially with the number of entangled qubits. Thus, although we can boost the task of automatically constructing the quantum circuits for many of the entanglement classes of 4 qubit states, in the end we will also face the wall of the exponential complexity of quantum entangled states with an arbitrary number of qubits.
	Nevertheless, the quantum circuits synthesized with machine algorithms with reinforcement learning can serve as a benchmark for more complex quantum compilers.

	\begin{acknowledgements}
We acknowledge support from the CAM/FEDER Project No.S2018/TCS-4342 (QUITEMAD-CM), Spanish MINECO grants MINECO/FEDER Projects, PGC2018-099169-B-I00 FIS2018, MCIN with funding from European Union NextGenerationEU (PRTR-C17.I1) an Ministry of Economic Affairs Quantum ENIA project. 
M. A. M.-D. has been partially supported by the U.S.Army Research Office through Grant No. W911NF-14-1-0103.
S.G. acknowledges support from a QUITEMAD grant.

We acknowledge the precious support of R. Fazio (ICTP and Università degli studi di Napoli "Federico II"), P. Lucignano (Università degli studi di Napoli "Federico II") and the Università degli studi di Napoli "Federico II". 

	\end{acknowledgements}

\appendix
 \section*{Appendix: Post-processing and normalization}
 \label{Appendix}
 \setcounter{equation}{0}
 \renewcommand{\theequation}{A.\arabic{equation}}
	As we mentioned in Sec.\ref{sub_II_A}, we have to check and fix the coefficients of the quantum states after we manage to approach them with the quantum circuits generated by our algorithm. In particular, we address a post-processing procedure that allow us to tune the coefficients in the superposition, if these latter do not meet the desired ones. This procedure consists in replacing one or more Hadamard or C-Hadamard gates of the circuit that results from the Q-learning procedure, with unitary gates, with one or more free parameters. In general, a unitary gate can be written as:
	\begin{small}
	\begin{gather}U(\theta,\phi,\lambda)=
		\bigg( {\begin{array}{*{20}{c}}
				\cos(\frac{\theta}{2})&-e^{i\lambda}\sin(\frac{\theta}{2})\\
				e^{i\phi}\sin(\frac{\theta}{2})&{e^{i(\phi+\lambda)}\cos(\frac{\theta}{2})}
		\end{array}} \bigg)
	\end{gather}
	\end{small}
	by setting the parameters $\theta$, $\phi$ and $\lambda$ we can build every single-qubit unitary gate. As our interest is to modify real coefficients, we need to use the $U$ gate as a pure rotation, in order to create an \textit{unbalanced} Hadamard gate (or C-Hadamard gate). Indeed, the Hadamard gate corresponds to $U(\frac{\pi}{2},0,\pi)$ and, leaving the $\theta$ parameter free, we have a pure rotation:
	\begin{small}
	\begin{gather}U(\theta,0,\pi)=
		\bigg( {\begin{array}{*{20}{c}}
				\cos(\frac{\theta}{2})&\sin(\frac{\theta}{2})\\
				\sin(\frac{\theta}{2})&{-\cos(\frac{\theta}{2})}
		\end{array}} \bigg)=U(\theta).
	\end{gather}
	\end{small}
	Notice that in the case of the C-U gate, its matrix representation reads:
	\begin{small}
	\begin{gather}\small C\text{-}U(\theta)=
		\begin{pmatrix}
			1&0&0&0\\
			0&1&0&0\\
			0&0&\cos(\frac{\theta}{2})&\sin(\frac{\theta}{2})\\
			0&0&\sin(\frac{\theta}{2})&{-\cos(\frac{\theta}{2})}
		\end{pmatrix}
	\end{gather}
	\end{small}
	We assume to replace, in the output circuit of the Q-learning algorithm, $m$ Hadamard or C-Hadamard gates, we call the i-th unitary gate $U_i(\theta_i)$ or $C\text{-}U_i(\theta_i)$, where $\theta_i$ is the i-th free parameter. We then apply to the initial state the circuit with the replacements, the resulting state after this quantum circuit has the same terms of the objective one in Eq.\eqref{eq:obj_state} but has still undetermined coefficients:
	\begin{gather}
		\ket{\Psi}_{out}=\sum_{j=1}^{16}f_j(\theta_0,\dots,\theta_m) \ket{\psi_j}.
	\end{gather}
	Here the coefficients $f_j(\theta_0,\dots,\theta_i,\dots,\theta_m)$ can assume any desired value, with the requirement of $\ket{\Psi}$ normalization.
	If the desired representative state of a SLOCC class reads:
	\begin{gather}
		\ket{\Psi}=\sum_{j=0}^{16}\alpha_j\ket{\psi_j},
	\end{gather}
	we can find the values of the parameters $\theta_i$ by solving the following equations system:
	\begin{small}
		\begin{gather}
			\left\{\begin{matrix}
				f_0(\theta_0,\dots,\theta_i,\dots,\theta_m)=\alpha_0 &&\\ 
				\vdots \\ 
				f_j(\theta_0,\dots,\theta_i,\dots,\theta_m)=\alpha_j && \; 	\;\text{where}\; \alpha_j\in \mathbb{R},\; \forall j ,\;\; \\
				\vdots \\
				f_n(\theta_0,\dots,\theta_i,\dots,\theta_m)=\alpha_n &&
			\end{matrix}\right.
		\end{gather}
	\end{small}
	On the left side we have the undetermined coefficients for each term, in form of trigonometric functions of $\theta_1,\dots,\theta_m$, where $m$ stands for the overall number of Hadamard and C-Hadamard gates replaced; on the right side we have the desired coefficients, $\alpha_0,\dots, \alpha_n$, where $n$ is the number of terms of the goal state. Despite the existence of the solution is not theoretically guaranteed for that system, in our cases we always find suitable values for $\theta_i$ that allow us to reach the representative state $\ket{\Psi}$ of the SLOCC class in exam.
	
	Let us take as an example the SLOCC class B1.1, belonging to the SLOCC family $L_{abc_2}$. With the Q-learning procedure we manage to find the circuit in Fig.~\ref{fig16:classB1.1circ}. Taking into account the normalization coefficients of the Hadamard and C-Hadamard gates, the resulting state after this circuit reads:
	\begin{small}
	\begin{gather}
	 \ket{\Psi}_{out}=\frac{1}{\sqrt{2}}\ket{0000}+\frac{1}{2}\ket{0110}+\frac{1}{2}\ket{1111}
	\end{gather}
	\end{small}
	in order to obtain a state that matches the normalized representative state of the class $\ket{\Psi}_{B1.1}=\frac{1}{\sqrt{3}}(\ket{0001}+\ket{0110}+\ket{1011})$ we can apply our post-processing procedure and replace the first Hadamard gate $H(C)$ with a unitary gate $U(C)$. The result reads:
	\begin{small}
	\begin{alignat}{2}
		\ket{\Psi}_{out}=\cos{\left(\frac{\theta_0}{2}\right)}\ket{0000}&+\frac{\sin{\left( \theta_0/2\right)}}{\sqrt{2}}\ket{0110}\\ \notag
		&+\frac{\sin{\left( \theta_0/2\right)}}{\sqrt{2}}\ket{1111}.
	\end{alignat}
	\end{small}
	It is straightforward that, in order to obtain $\ket{\Psi}_{out}=\ket{\Psi}_{B1.1}$the system to solve is the following:
	\begin{gather}
		\left\{\begin{matrix}
				\cos{\left(\frac{\theta_0}{2}\right)} = \frac{1}{\sqrt{3}} && \\
				&&\\
				\frac{\sin{\left( \theta_0/2\right)}}{\sqrt{2}}=\frac{1}{\sqrt{3}} && \\
				&&\\
				\frac{\sin{\left(\theta_0/2\right)}}{\sqrt{2}} = 	\frac{1}{\sqrt{3}} &&
		\end{matrix}\right.
	\end{gather}
	which solution include:
	\begin{gather}
		\left\{\theta_0 \to 2\arctan \left(\sqrt{2}\right)+4 \pi k \; \middle|\; k\in \mathbb{Z}\right\}.
	\end{gather}
	The resulting state with this choice of the $\theta_0$ parameter, is the desired one. Notice that in this case, as we replaced a single Hadamard gate, we have only one unknown. In some cases we need to replace more than one Hadamard or C-Hadamard gate to obtain a solvable system.

\end{document}